\documentclass[twocolumn]{pasj00}
\renewcommand{\tabcolsep}{4pt}

\begin{document}
\SetRunningHead{T. Tsuji \& T. Nakajima}{Near Infrared Spectroscopy of M Dwarfs. I. CO Molecule as an Abundance Indicator of Carbon }

\title{ Near Infrared Spectroscopy of M Dwarfs. I.  

CO Molecule as an Abundance Indicator of Carbon \thanks{Based on data 
collected at Subaru Telescope, which is operated by the National Astronomical 
Observatory of Japan.} }

\author{Takashi \textsc{Tsuji}} %
\affil{Institute of Astronomy, The University of Tokyo,
2-21-1 Osawa, Mitaka-shi, Tokyo, 181-0015}
\email{ttsuji@ioa.s.u-tokyo.ac.jp }

\and
\author{Tadashi \textsc{Nakajima}}
\affil{National Astronomical Observatory of Japan, 2-21-1 Osawa,
Mitaka-shi, Tokyo, 181-8588}
\email{tadashi.nakajima@nao.ac.jp}


\maketitle

\begin{abstract}

Based  on the near infrared  spectra of 42 M dwarfs,
carbon abundances are determined from the ro-vibrational lines 
of CO 2-0 band.  We apply  $T_{\rm eff}$ values based on the angular 
diameters (15 objects) and on the infrared flux method (2 objects)
or apply a simple new method
using a  {\rm log}\,$T_{\rm eff}$ (by the angular diameters and by the
infrared flux method) -- $M_{\rm 3.4}$ (the absolute magnitude at 
3.4 $\mu$m based on the $WISE~W1$ flux and the Hipparcos parallax) 
relation to estimate $T_{\rm eff}$ values of objects for which angular 
diameters are unknown (25 objects). 
Also, we discuss briefly the HR diagram of low mass stars.

On the observed spectrum of the M dwarf, the continuum is depressed 
by the numerous weak lines of H$_2$O and only the depressed continuum or
the pseudo-continuum can be seen. On the theoretical spectrum of the M dwarf,
the true continuum can be evaluated easily but the pseudo-continuum  
can also be evaluated  accurately thanks to the recent H$_2$O line database.
Then spectroscopic analysis of the M dwarf can be done by 
referring to the pseudo-continua both on the observed and theoretical
spectra.  Since the basic principle of the
spectroscopic analysis should be the same whether the true- or 
pseudo-continuum is referred to, the difficulty related to the continuum
in cool stars can in principle be overcome. Although this procedure
can easily be applied to the spectral synthesis method, we propose a simple
method of analyzing EWs by taking the contamination of H$_2$O
on CO lines into account: We measure the EWs of the CO blends (i.e.,
not necessarily limited to a single line) including H$_2$O as contamination
and analyze them by the theoretical EWs evaluated from the synthetic spectrum
including the effect of H$_2$O contamination as well.

In dense and cool photospheres of M dwarfs, almost all the carbon atoms are 
in stable CO molecules which remain almost unchanged for the changes of the
physical condition in the photospheres. For this reason, the numerous CO 
lines can be excellent abundance indicators of carbon, and carbon 
abundances in late-type stars  can best be determined in M dwarfs rather 
than in solar type stars. This somewhat unexpected advantage of M dwarfs in 
abundance analysis has not neessarily been recognized very well so far,
but we determine carbon abundances rather easily to be log\,C/H between -3.9 
and -3.1 in 42 M dwarfs. The resulting C/Fe ratios for most M dwarfs are 
nearly constant at about the solar value based on the classical high carbon 
abundance rather than on the recently revised lower value. This result 
implies that the solar carbon abundance is atypical for its metallicity 
among the stellar objects in the solar neighborhood if the downward revised 
carbon abundance is correct.  

\end{abstract}

\section{Introduction}
  M dwarf stars are the largest stellar group in the solar
neighborhood. Also, because of their small masses, the lifetimes of M 
dwarfs are  longer than the age of the Universe, and M dwarfs 
are composed of the samples of all the populations including
very old ones.  Unfortunately our understanding on this  large and rich 
stellar group is rather poor compared to other stellar groups. 
This should largely be due to difficulty of observing M dwarfs because of 
their faintness, and the difficulty is most severe for the M dwarfs of the old
population in the distant halo. 
Recent progress, however, overcame the observational
barriers to some extent, especially in the field of stellar interferometry, 
infrared spectroscopy using new infrared detectors, astrometry and photometry
from space, etc.

In addition to the faintness of M dwarfs,
spectroscopic analysis of M dwarfs has also been hampered by the
heavy blending of many atomic and molecular lines, especially in the
optical region. This difficulty can be reduced somewhat in the
infrared region, as has been shown by \citet{Mou78} who has
carried out abundance analysis based on the  near infrared spectra of 
six M dwarfs observed with the
Fourier Transform Spectrometer (FTS) at KPNO. After this promising
attempt in the 1970's, however, infrared spectroscopy of M dwarfs 
at high resolution could not be pursued further, because 
infrared detectors were  not yet sensitive enough for faint M dwarfs.  
Recent progress of the infrared two-dimensional detectors finally made it 
possible to pursue spectroscopic analysis of M dwarfs further,
as has already been demonstrated by \citet{One12} who have analyzed
the infrared high resolution spectra in the $J$ band region of
 about a dozen of M dwarfs.

Although spectroscopic analysis of the M dwarf star  has been deemed 
difficult in general, we have some reasons why we  study such
a difficult case. For example, carbon and oxygen abundances in
late-type stars are generally difficult to determine and their abundances 
are still controversial even for the Sun. This is because the abundance 
indicators of carbon such as the high excitation lines of neutral
carbon and molecular lines such as of C$_2$ and CH (e.g., \cite{Gre91};
\cite{Asp05}), for example, are mostly pretty model sensitive. In
contrast, such a difficulty can be reduced in M dwarfs for which numerous
lines of stable molecules such as CO and H$_2$O can be used as abundance 
indicators of carbon and oxygen, respectively. Such an advantage in
the spectroscopic analysis of M dwarfs has not necessarily been well
recognized so far and we hope to explore such a  possibility in some
detail in this paper.
  
Also, recent interests in M dwarfs come from  possibilities to detect
Earth-type planets around M dwarfs and to know the nature of such
M dwarfs that host planets. Until the present, it has been suggested that 
metallicity of the planet hosting M dwarfs may be high (e.g., \cite{Joh09}). 
Also habitability of the planets may depend largely on their chemical
environment and accurate abundance determination of the parent
M dwarfs should be an important issue in investigating habitable planets
themselves. However, direct  abundance determinations in M dwarfs based on 
detailed spectroscopic analysis are still limited,
and most estimations of metallicity were based on indirect 
methods based on low resolution spectroscopic or photometric indexes
(e.g., \cite{Roj12}; \cite{Nev13}). 

For extending detailed spectroscopic analysis to M dwarfs, however,
another difficulty is the heavy veiling of the spectra
by molecular bands, which obscures the true continuum needed to
measure accurate line strengths (or equivalent widths). 
If we apply the spectral synthesis (SS) method, which
has been more popular in recent years, the situation is the same,
since the synthetic spectrum has usually been evaluated by referring to
the true-continuum. Given that the problem of the continuum  
is a common feature that cannot be avoided in cool stars, we think that
it is useful to examine the nature of the veil opacities in some details and 
reconsider such an analysis carried out by referring to the  true-continuum.
In fact, it is now possible to evaluate the pseudo-continuum
level fairly accurately by virtue of the  recent progress in molecular
line database  such as  HITEMP2010 (\cite{Rot10}) and many references
cited therein (e.g., \cite{Bar06}). Then,  we may abandon to refer to the 
true-continuum and reformulate the spectroscopic analysis in such a way
as to proceed by referring to the pseudo-continuum defined by the
molecular veil opacities.  
The basic principle of the spectroscopic analysis should be the same 
whether the true- or pseudo-continuum is referred to.

Now, the observational barriers are being overcome to some extent and 
the method of analysis can hopefully be improved somewhat as outlined above, 
we hope to pursue the possibility of determining  accurate
abundances directly from the observed spectra of M dwarfs. 
 
In this paper, we first introduce our observed data (section\,2) and then
examine the necessary data and tools for spectroscopic analysis such as 
fundamental stellar parameters, model photospheres,
and molecular data (section\,3). We then carry out a preliminary
analysis based on the conventional method disregarding the background
opacity due to numerous H$_2$O lines (section\,4), and consider the effect
of H$_2$O contamination consistently in the analysis  referring to 
the pseudo-continuum (section\,5). Finally, we discuss some topics including
method of analysis,  carbon abundances in M dwarfs, accuracy of 
abundance analysis, and HR diagram at the end of the main sequence
 (section\,6).

\section{Observations}

We observed 42 M dwarfs listed in the table 1,  which
were selected based on at least one of the following three criteria.

\vskip 3mm 
\noindent
1. The distance of the target ($d < $ 30 pc) is known by  
parallax measurements mostly by Hipparcos \citep{Lee07}.  All the target 
stars satisfy this criterion.

\noindent
2.  The radius of the target star was obtained by interferometry
(\cite{Boy12}, references therein). 16 stars whose names are marked by 
$\dagger$ in table\,1  belong to this case.

\noindent
3. The target star hosts a planet (or planets) (\cite{Mar98}; \cite{Del98};
\cite{But04}; \cite{Bon05}; \cite{But06}; \cite{For09}; \cite{Joh10}; 
\cite{Hag10}; \cite{App10}; \cite{How10}; \cite{For11}).
Ten stars whose names are marked by $\ddagger$ in table\,1 belong 
to this case.

\vskip 3mm

Observations were carried out at the Subaru Telescope on 2013 May 9 and 
November 16 UT using the echelle mode of the 
Infrared Camera and Spectrograph (IRCS; \cite{Kob00}) with natural guide
star adaptive optics. The slit width of 0.14$^{\prime\prime}$ was sampled 
at 55 mas pixel$^{-1}$, and the resolution was about 20000 at $K$.
The echelle setting was ``$K^+$'', which covered about a half of the $K$ 
window with the  orders,  wavelength segments and pixel-wise 
dispersions given in table 2.
The targets were nodded along the slit, and observations were taken in ABBA 
sequence, where A and B stand for the first and second positions on the slit. 
Total exposure time ranged from 24 seconds (GJ 820 A) to 54 minutes
(GJ 3348 B). The night in May was photometric, while that in November was 
spectroscopic.  Signal-to-noise ratios of reduced spectra at around 
23000{\AA}  given in the fifth column of table\,1,
were typically higher for the run in May.  At the beginning of the night
in the May run  and at the ending of the night in the November run,
a rapidly rotating B8 star Regulus ($\alpha$ Leo) was observed as the 
calibrator of telluric transmission.

During the target acquisition by $K$ band imaging, GJ 797 B was spatially 
resolved into an equal binary with a separation of 0.3$^{\prime\prime}$
with NE and SW components. By placing the slit perpendicularly to the 
separation, each component was observed. 

Data reduction was carried out using the standard IRAF\footnote{IRAF is 
distributed by the National Optical Astronomy Observatory, which is
operated by the Association of Universities for Research in Astronomy, Inc., 
under cooperative agreement with the National Science Foundation.}
  routines in the \texttt{imred} and \texttt{echelle} packages. 
After extraction of one dimensional spectra,
wavelength calibrations were calculated using telluric absorption lines in 
the spectra of Regulus. 
After wavelength calibrations of one dimensional spectra of A and B positions,
they were coadded to produce combined spectra. 
The combined spectra were normalized by the pseudo-continuum levels and then 
calibrated for telluric absorption 
using the spectra of Regulus.

\vspace{2mm}

-----------------------------------------------------

table\,1: Target stars (p.24).

table\,2: Echelle setting (p.25).

-----------------------------------------------------
  
\section{Preparations  for Spectroscopic Analysis  }
The basic parameters that characterize stellar spectra are
effective temperature, surface gravity, chemical composition, and
micro-turbulent velocity. However, it is difficult to determine
all these parameters from the observed spectra, and it is helpful
to obtain some of these parameters from other sources. We survey
such a possibility especially in the recent stellar interferometry and
in the databases based on space observations (subsection 3.1).  With the 
basic parameters
at hand, we generate model photospheres needed to compute theoretical spectra 
by which to analyze the observed spectra, and we briefly discuss
our models (subsection 3.2). In the computation of the theoretical spectra,
accurate molecular data are indispensable and we discuss the molecular 
data we are to use (subsection 3.3). 

\subsection{Fundamental Parameters of M Dwarfs}

Thanks to the recent interferometric measurements of the angular diameters
of M dwarfs, effective temperatures of dozens of M dwarfs are now known
with sufficient accuracy (e.g., \cite{Boy12}) and about a third of our
sample belong to this case.
The problem is how to estimate effective temperatures of
M dwarfs for which angular diameters are not known..
 One possibility may be to apply photometric 
data, and the best way for this purpose may be to apply
 the infrared flux method \citep{Bla80}.
 We  applied this method by the use of the $L^{'}$ band flux  
but only to a limited number of M dwarfs \citep{Tsu96}. 
For extending this method to our M dwarfs,  
we looked for the $L^{'}$ band data in the literature, but we could not
find them in the ground-based photometry for most of our sample.
But we noticed  that the $WISE ~ W1$ band flux centered at 3.4\,$\mu$m 
\citep{Wri10} can be used for this purpose.

  Meanwhile, however, we find it possible to use the $WISE$ data not
in the infrared flux method but in another way: Namely, for M dwarfs
for which interferometric $T_{\rm eff}$ values are available, we convert
the $W1$ flux to the absolute magnitude at 3.4\,$\mu$m, $M_{\rm 3.4}$, by 
the use of the Hipparcos parallax \citep{Lee07}. Then we use the
resulting $M_{3.4} - {\rm log}\,T_{\rm eff}$ relation to infer $T_{\rm eff}$
from $M_{3.4}$ which is available for all the M dwarfs we are to analyze.
There are 33 K and M dwarfs for which interferometrically determined 
$T_{\rm eff}$  are known with precision 
to better than 5\% in table 6 of \citet{Boy12}. However, the $WISE$ fluxes
were not resolved for the binary components  GJ\,338A, GJ\,338B,
GJ\,702A, GJ\,702B, GJ\,820A, and GJ\,820B, and we use the remaining 
27 stars. To include later M dwarfs,
we also consider nine M dwarfs whose $T_{\rm eff}$ values are
determined by the infrared flux method \citep{Tsu96}. These 27 and 9 data are
 summarized in tables\,3 and 4, respectively. For these 36 objects with
 known $T_{\rm eff}$,  $M_{\rm 3.4}$ values  are plotted against 
log\,$T_{\rm eff}$ values in figure 1.  Except for a few cases, 
$M_{\rm 3.4}$ and log\,$T_{\rm eff}$  follow reasonably tight 
correlation and we draw a mean curve shown by the dashed line in figure 1.

The resulting $M_{\rm 3.4}$ - log\,$T_{\rm eff}$ relation
is used to estimate $T_{\rm eff}$ of M dwarfs for which $T_{\rm eff}$ 
values are not known. The resulting $T_{\rm eff}$ values
together with $M_{\rm 3.4}$ values are given in table\,5 for 25 M dwarfs.
For two objects, GJ\,273 and GJ\,725B, which deviate from the mean relation
in figure\,1, we also estimate their $T_{\rm eff}$ values by using the 
$M_{\rm 3.4}$ - log\,$T_{\rm eff}$ relation and the results are included
in table\,5. The resulting $T_{\rm eff}$ = 3415\,K by the $M_{\rm 3.4}$-method
is higher compared with $T_{\rm eff}$ = 3150\,K (table 4) by the infrared flux 
method  for GJ\,273, while  $T_{\rm eff}$ = 3337\,K by the 
$M_{\rm 3.4}$-method is higher compared with the  $T_{\rm eff}$ =
3104\,K (table 3) by the interferometry  for GJ\,725B. 
We should in principle respect the results by the direct observations,
but the consistency among all the data should  be considered at the same time.
We analyze the spectra of these two M dwarfs by the two different
$T_{\rm eff}$ values - the high and low, and examine the overall consistency
at the end (subsection\,6.2). We attach H and L to the star names to 
discriminate the cases of high and low $T_{\rm eff}$, 
e.g., GJ\,273-H and GJ\,273-L.

In applying the $M_{\rm 3.4}$-method, the accuracy of the
input data, especially of the $WISE$ photometry is most crucial. 
Generally, the major problem in the photometry of relatively bright
objects is the possible saturation of the detectors.
In this regard, the $WISE$  data quality has been examined in detail 
by \citet{Ave12} who showed that the $W1$ band flux centered at 3.4\,$\mu$m 
is of high quality and quite reliable. This is in marked contrast to
the $W2$ band flux centered at 4.6\,$\mu$m, which was showed to be not
free from saturation effect and will result in significant errors for 
bright stars. By the way, \citet{Ave12} showed that there is no 
mid-infrared excess in about a hundred M dwarfs by the analysis of the
$WISE$ data and concluded that there
is no clear evidence for a debris disk around M dwarfs. This result
implies that the visibilities measured by the interferometers are  not
likely to be disturbed by emission and/or scattering by the circumstellar
dust.  However, detailed analysis should be needed on deviating stars
in figure\,1 to see if  such a conclusion can be applied to them.
 
We also plot $T_{\rm eff}$ values by  different
methods against the spectral types in figure\,2. For this purpose, we 
find that 29 M dwarfs in our sample are included in 426 M dwarfs classified 
on a uniform system by \citet{Joy74} who used the strengths of TiO bands 
in the blue. As for remaining 13 M dwarfs, we apply the spectral types
found in SIMBAD. Although these types due to different authors are
naturally not on a uniform system, we confirm in our sample of 29 M
dwarfs noted above that the SIMBAD types agree with those by
\citet{Joy74} within 1 subtype for the 28 out of 29 M dwarfs and within 
1.5 subtype for a remaining object GJ\,205.  
Inspection of figure\,2 reveals that the spectral type - $T_{\rm eff}$
relation is not defined very well, but most stars follow
a mean relation (dashed line) with dispersion of $\pm$ 1 subtype
(dotted lines)\footnote{A large deviating star GJ\,205 is very
metal-rich and may be classified to be too late for its $T_{\rm eff}$. 
GJ\,611B is metal poor and may be classified to be too early for its 
$T_{\rm eff}$.  But we have no explanation for GJ\,273-L, GJ\,686, and 
GJ\,725B-L.}. It is to be noted that there is little systematic effect
among the $T_{\rm eff}$ values determined by the interferometry (filled 
circles), the infrared flux method (open circles), and the 
$M_{\rm 3.4}$-method (filled triangles). 

The radii based on the angular diameters and masses mostly based on
the mass-luminosity relations are also reproduced in 8- and 9-th
columns, respectively, in table\,3  from \citet{Boy12}. The log\,$g$ values
based on these radii and masses are given in the 10-th column of table\,3,  
but only for those objects which are analyzed in the present
paper. For the objects for which $T_{\rm eff}$ values are determined
by the infrared flux method and by the $M_{\rm 3.4}$ - method, radii and 
masses are estimated by applying the empirical equations (8) and (10),
respectively, given by  \citet{Boy12}. The resulting values of $R/R_{\odot}$,
$M/M_{\odot}$, and log\,$g$ are given in the last three columns
in tables\,4 and 5 (but in table\,4,  only for
 objects which are  analyzed in the present paper).
Now, $T_{\rm eff}$ and  log\,$g$ values are ready for the  42 objects we
are  to analyze in this paper, and these data are plotted in figure\,3
together with the result from the evolutionary models by \citet{Bar98}. 
Note that GJ\,273 and GJ\,725B appear twice for the different sources
of $T_{\rm eff}$ values noted above. Also, GJ\,273-L and, especially 
GJ\,725B-L, show large deviations from other objects in figure 3.

\vspace{2mm}
-------------------------------------------------------------------

figure\,1: $M_{\rm 3.4}$ vs. log\,$T_{\rm eff}$ (p.15).

figure\,2: $T_{\rm eff} $  vs. Sp. types (p.16).

figure\,3: log\,$g$ vs. $T_{\rm eff}$ (p.16).

\vspace{2mm}

table\,3: $T_{\rm eff}$ by the interferometry (p.25).

table\,4: $T_{\rm eff}$ by the infrared flux method (p.26).

table\,5: $T_{\rm eff}$ by $M_{\rm 3.4}$ -- log\,$T_{\rm eff}$ (p.27).

-------------------------------------------------------------------

\subsection{Model Photospheres of M Dwarfs}

We apply the model photospheres of M dwarfs included as a part
of  our Unified Cloudy Models (UCM) (\cite{Tsu02}; \yearcite{Tsu05}). 
Our present M dwarf sample is
not so cool as to form dust clouds in their photospheres
and we apply our dust free models for two abundance cases: {\it case a}
(Ca-series) with the high C \& O abundances (log\,$A_{\rm C} = -3.40$ \&
log\,$A_{\rm O} = -3.08$\footnote{$A_{\rm C}$ and  $A_{\rm O}$ are the
number densities of carbon and oxygen, respectively, relative to hydrogen. 
The abundances for {\it case a} including 34 elements are given in Table\,1 of 
\citet{Tsu02}, and those  for {\it case c} are the same as  {\it case a} 
except for $A_{\rm C}$ and  $A_{\rm O}$.}) and {\it case c} (Cc-series) with 
the low C \& O abundances (log\,$A_{\rm C} = -3.61$ \& log\,$A_{\rm O} = 
-3.31$), which are
based on the classical solar abundance (\cite{And89}; \cite{Gre91}) and
a downward revised solar one \citep{All02}, respectively. Our grid 
covers $ T_{\rm eff}$ between 700 and 4000K with a step of 100\,K for
log\,$g$ = 4.5, 5.0, and 5.5. However, inspection of tables\, 3 $\sim$ 5 
reveals that log\,$g$'s of most M dwarfs are somewhere between log\,$g$ = 4.5
and 5.0. For this reason, we  generate an additional series of 
models with log\,$g$ = 4.75 ($T_{\rm eff} = 2600 \sim 4000$ K) in our
UCM grid, and this series of models are used in the
first iteration of our preliminary analysis (section\,4). 

However, we  find that the abundance analysis is sensitive to the 
fundamental parameters, especially to log\,$g$ (subsection\,4.3).
Then we finally decide not to use the models from our grid but to generate
a specified model for $ T_{\rm eff}$ and log\,$g$ of each M dwarf,
and use it in and after the second iterations.  
If the carbon abundance by our first iteration of the  preliminary
analysis appears to be closer to the classical 
solar abundance (or to the recent solar abundance), we apply the {\it case a} 
(or {\it case c}) abundance as the input C \& O abundances for our models   
\footnote{We designate our model by cloud type/abundance case/$T_{\rm eff}$/
log\,$g$. For example, a model of no dust cloud (or clear model) with 
{\it case a} abundance, $T_{\rm eff}$ = 3360\,K, and log\,$g$ = 4.85 is 
referred to as Ca3360c485. As for details about our models of M, L, and T 
dwarfs, see an updated UCM database at 
http://www.mtk.ioa.s.u-tokyo.ac.jp/$\sim$ttsuji/export/ucm2. }.

\subsection{Molecular Data}

 The spectroscopic data of CO are relatively well known.  Our
 CO data are based on  the line position data \citep{Gue83} and 
intensity data (\cite{Cha83}).  Under the high density of the 
photospheres of M dwarfs, another important issue
is the collision broadening, which depends on the physical properties of the 
interacting atoms or molecules. The collision half-width $\gamma$ is
relatively well studied for the collision partners in the air
 (N$_2$, O$_2$...) at room temperature, but unfortunately not for collision 
partners in stellar photospheres (H, He, H$_2$...) at elevated temperatures.
Usually, collision half-width $ \gamma$ is represented by
    $$ \gamma = \gamma_{0}{p \over p_{0} }( {T_{0} \over T}  )^{n} , 
\eqno (1)    $$
where $\gamma_{0}$ is the collision half-width measured at a reference
temperature $T_{0}$ (e.g. 296\,K) and gas pressure $p_{0}$.  The CO 
air-broadened  values of $\gamma_{0}$ can be found in the CO database 
included in HITEMP2010 (\cite{Rot10}), but those by molecular hydrogen 
appropriate for  temperatures as high as 3000\,K are
available only for CO pure rotation transitions  (\cite{Fau13})
so far as we are aware. As an example of the effect of the collision partner,
we compare the H$_2$ broadening half-widths (converting from MHz/Torr
to cm$^{-1}$/atm unit from \citet{Fau13}) and air-broadening ones 
(\cite{Rot10}) for the case of CO pure rotation transitions in table\,6.

Inspection of table\,6 reveals that the effect of collision partners is
rather small in the low $J$ transitions but quite appreciable in the
high $J$ transitions. The values of $\gamma_{0}$ for the 
air-broadening of CO 2-0 vibration-rotation transitions which we analyze 
in this paper are rather similar (\cite{Rot10}) to the case of the pure 
rotation transitions shown in table\,6. Although it is possible that the 
H$_2$-broadening is about 50 \% larger than the air-broadening, we
apply the medium value of $\gamma_{0} = 0.05$\,cm$^{-1}$/atm for the 
air-broadening data \citep{Rot10} to all the CO 2-0 lines 
we study in this paper ($J \approx 30 - 70$). We hope that measurements of the 
H$_2$-broadening at  high temperatures can be extended to the 
vibration-rotation transitions of CO in the near future. 
  
For the reason to be noted in subsection\,4.1, we decide to concentrate
to the CO lines near the (2,0) bandhead region and  select CO blends
listed in table\,7 in which the spectroscopic data of CO lines are given.
We hoped that  this spectral region may be relatively
free from other  molecular bands.  In fact, our line-list includes
some lines of CN Red System, ro-vibrational lines of OH, HF, H$_2$ etc.,
but none of these molecular lines is prominent in the region we are to study.
However, we notice that the wings of H$_2$O $\nu_{1}$ and $\nu_{3}$ bands 
centered at 2.7\,$\mu$m and $\nu_{2} + \nu_{3}$ band centered at 1.87\,$\mu$m
 extend to the region of the CO 2-0 band and 
numerous weak H$_2$O lines produce appreciable
effect  on the continuum level. Also some CO lines are blended with the H$_2$O
lines in such a way that their equivalent widths are modified by the blends.

We first examine the effect of H$_2$O lines with the use of the line-list
by \citet{Par97} on a model of Cc3500c50
and $\xi_{\rm micro}$ = 1\,km\,sec$^{-1}$. 
The result is shown in figure\,4a in which the predicted spectrum with 
the resolution of about $10^{5}$ is shown by the thin line and that
convolved with the slit-function (Gaussian) with FWHM = 16\,km sec$^{-1}$
(the resolution of our observed spectra) by thick line. 
Next, we do the same but with the use of a new water line-list in HITEMP2010
(\cite{Rot10}) which has extensively used the BT2 database by
\citet{Bar06} (figure\,4b). Although the general patterns of the H$_2$O
spectra in figures\,4a and 4b
are not very different, the depressions of the continua are much larger in 
figure\,4b (more than 3\,\%) than in figure\,4a (less than 1\,\%).
For comparison, the predicted spectrum of CO alone shows the
continuum clearly (figure\,4c).  Although the collision half-widths of 
H$_2$O lines span the larger range compared with those of CO, we apply 
the same mean value as for CO for   
H$_2$O lines in which the damping wings are not yet developed.
 
The number of H$_2$O lines with cut-off at the integrated intensity
\footnote{line absorption cross-section[cm$^2$] integrated in the wavenumber
[cm$^{-1}$] space over the entire line profile.} 
of $S(T = 2500\,{\rm K}) \approx 3 \times 10^{-27}$\,cm\,molecule$^{-1}$
in the spectral region shown in figure\,4 (22925 - 23020\,{\AA}) is 3021 
by the line-list of \citet{Par97} while it is 32970 by BT2-HITEMP2010. 
The number of lines in the original BT2-HITEMP2010 
database is still several times larger. This large increase is due to 
inclusion of many weak lines in BT2-HITEMP2010, and it is such numerous 
weak lines that play important role in depressing the continuum 
level noted in figure\,4b. Also, with such a high line density of  
about 350 lines per 1\,{\AA}  interval for the cut-off noted above,
the H$_2$O opacity should be  smeared out and the pseudo-continuum
can be well defined at nearly constant level as shown in figure\,4b. 
For this reason, spectral analysis can be carried out by referring
to the pseudo-continuum instead of the true-continuum, as will be
discussed in section\,5.

\vspace{2mm}
-------------------------------------------------------------------

figure\,4a,b,c: Theoretical spectra of H$_2$O and CO (p.17).  

\vspace{2mm}

table\,6: Effect of collision partners on $\gamma$ (p.27).

table\,7: Spectroscopic data of the CO lines (p.28). 

-------------------------------------------------------------------

\section{A Preliminary Analysis} 
  We first examine the CO spectra (subsection\, 4.1) and apply a usual method
 of abundance analysis to the CO lines  without considering the possible 
effect of the blending of the H$_2$O lines (subsection\,4.2). 
The external errors due to the uncertainties in the fundamental 
parameters including the overall metallicity are discussed (subsection\,4.3). 

\subsection{CO Lines}
  In this paper , we focus our attention to the lines of the CO
first-overtone band. Although
CO lines are clearly observed in the spectral region observed,
 most lines are blends of two or more CO lines in our medium
resolution spectra of $R\approx 20000$ (figure\,4c). 
Also, we notice that H$_2$O lines are already disturbing the region
of the CO first overtone band.  As can be inferred from figure\,4b, numerous 
weak H$_2$O lines
depress the continuum level on one hand and some relatively strong H$_2$O lines
disturb the individual CO lines as blends on the other. 
All these facts make it difficult to measure equivalent widths accurately
and, in particular, it is very difficult to measure weak lines.

We first hoped to measure as many CO lines as possible recorded in 
our echelle spectra since it is essential to include weak unsaturated 
lines and saturated stronger lines together to determine
micro-turbulent velocity, which is the important parameter next to the
effective temperature and gravity in the interpretation of stellar spectra.
 However, the situation noted above, 
namely blending of CO lines themselves, blending
of H$_2$O lines disturbing both continuum level and CO line strength,
and difficulty to measure weak CO lines, makes it difficult to determine
micro-turbulent velocity spectroscopically. On the other hand, the
micro-turbulent velocity in the photosphere of M dwarfs may be less than
1 km sec$^{-1}$ ( e.g., \cite{Bea06}) while the thermal velocity of CO is 
1.33 km sec$^{-1}$ for $T \approx 3000$\,K. Then the role of the
micro-turbulence in the line-broadening in the photospheres of M dwarfs
may not be so important as in M giant stars in which the micro-turbulent
velocity is larger than 1 km sec$^{-1}$. For this reason, we give up
to determine the micro-turbulent velocity spectroscopically in our M dwarfs 
and assume it to be 1 km sec$^{-1}$ throughout this paper.  

Once we give up to determine the micro-turbulent velocity, it is almost
useless to measure many CO lines which are more or less disturbed by
various noises, and we focus our attention  to the CO (2-0) bandhead
region shown in figure\,4c. In this region, the effects of H$_2$O $\nu_{1}$
and $\nu_{3}$ bands are relatively small compared to the longer
wavelength region and there is no overlapping  of various CO bands. 
We then concentrate to determine carbon abundance from the restricted CO 
features listed in table\,7.

\subsection{Analysis by Blend-by-Blend }
  All the  CO lines we are to analyze are blends of two CO lines on our
medium resolution spectra (figure\,4c). In such a case, the usual 
curve-of-growth method cannot be applied. Instead, spectral synthesis (SS) 
method may be considered, since it is possible to treat the blended 
features easily by this method.
However, SS method has been applied so far only when the true-continuum
could have been defined and, for this reason, it may be difficult to
apply to M dwarfs for which the true-continuum cannot be defined in general.
Although we hope to relax this difficulty as will be detailed in section\,5, 
we here extend the line-by-line (LL) method extensively applied to the
unblended lines in red giant stars (e.g., \cite{Tsu08}) to the case of 
blended lines in M dwarfs and now the method should better be referred
to as the blend-by-blend (BB) method. 

   We measure the equivalent widths (EWs) of the blends composed of two CO
lines listed in table\,7, and the results are given in table\,14\footnote{
This  table is available only in the electronic version as a supplementary 
data. Note that the line of Ref. no.\,11 is not measured because this line 
is blended with Sc\,I line and hence not used in our analysis.}.
Now the problem is to determine the carbon abundance from these EWs.
Since almost all the carbon atoms are in CO molecules in the photosphere 
of M dwarfs, the carbon  abundance is almost the same as the CO abundance.
However, it is not possible to obtain CO and hence carbon abundances directly 
from the measured EWs.  But the reverse is possible; i.e.,  determine directly
the EW of a CO blend for a given carbon abundance. Then, we start from the
carbon abundance log\,$A_{\rm C}^{(0)}$ of the model photosphere used 
(i.e., log\,$A_{\rm C}^{(0)} =-3.61$  for our models of Cc-series) and 
compute the EW of a CO blend. If the resulting EW is smaller than the 
observed EW, the assumed carbon abundance should be too small. We then 
compute EWs, $W(\delta)$, for the logarithmic abundance corrections of 
$\Delta\,{\rm log}\,A_{\rm C} = \delta$ = +0.3 and +0.6, and we have so to 
speak a mini curve-of-growth for the CO blends defined by 
log\,$W(\delta)/\lambda$ vs. $\delta$ ($\delta$ = 0.0, 0.3, and 0.6).
With this mini curve-of-growth, the abundance correction to the 
initial value of log\,$A_{\rm C}^{(0)} =-3.61$ can be obtained from
the observed EW of the CO blend. This process is repeated to all the
CO blends listed in table\,7 except for the lines of Ref. nos.\,13 and 14 
which appear to be badly blended with H$_2$O lines, and the mean of the 
resulting abundance corrections for about a dozen of the CO blends is
 obtained.

      An example of the above noted procedures is shown in figure\,5a
for the case of GJ\,412A:
The ordinate is the logarithmic abundance correction obtained
from the observed equivalent width indicated by the abscissa. 
The resulting mean correction $\Delta{\rm log}\,A_{\rm C}^{(1)}$
= -0.18 is shown by the dashed line, and the result indicates that the
carbon abundance should be lower in GJ\,412A than in the value assumed.
The revised logarithmic carbon abundance is obtained by
$$ {\rm log}\,A_{\rm C}^{(1)} =  {\rm log}\,A_{\rm C}^{(0)} +  
\Delta{\rm log}\,A_{\rm C}^{(1)}.      \eqno(2)$$ 
Another example  is shown in figure\,6a for HIP57050 which appears to be very 
carbon rich (about 3 times larger compared to the initial solar value).
Also, the case of the coolest object GJ\,406 is shown in figure\,7a.
  The results for our 42 M dwarfs are summarized in table\,8   in which
object identification, $T_{\rm eff}$, log\,$g$, model photosphere used in 
the first iteration, the logarithmic carbon abundance correction 
$\Delta{\rm log}\,A_{\rm C}^{(1)} $ , and the resulting logarithmic carbon 
abundance log\,$A_{\rm C}^{(1)}$  are given through first to 6-th columns. 

    Now we have a rough idea on the carbon abundance for each M dwarf
and we proceed to the second iteration, in which we apply our models
of Ca or Cc series 
depending on whether ${\rm log}\,A_{\rm C}^{(1)} \gtrsim -3.50$ or
 $ < -3.50$.  We also use the model specified for $T_{\rm eff}$ 
and log $g$ of each object instead of the model from our UCM grid. 
We then assume the carbon abundance resulted from the
first iteration as an input value and the same procedure as for the first
iteration is repeated. Some examples of this process are shown in
figures\,5b, 6b, and 7b for GJ\,412A, HIP\,57050, and GJ\,406, respectively.
The resulting second  logarithmic abundance corrections shown by the dashed
lines in figures\,5b, 6b, and 7b are  still non-zero but rather small in
all the three cases.
The results for 42 M dwarfs are summarized in table\,8, in which the specified
model photosphere used, the second logarithmic abundance correction  
$\Delta\,{\rm log}\,A_{\rm C}^{(2)}$, and the resulting logarithmic
carbon abundance
$$ {\rm log}\,A_{\rm C}^{(2)} =  {\rm log}\,A_{\rm C}^{(1)} +  
 \Delta{\rm log}\,A_{\rm C}^{(2)},    \eqno(3) $$ 
are given in 7-th, 8-th, and  9-th columns, respectively. 

Finally, we repeat the same procedure starting from the new logarithmic carbon
abundance $ {\rm log}\,A_{\rm C}^{(2)} $
and confirm that the resulting correction is almost null 
as shown in figures\,5c, 6c, and 7c for GJ\,412A, HIP\,57050, and GJ\,406,
respectively,
confirming  that  our preliminary analysis by the BB method has converged. 

\vspace{2mm}
----------------------------------------------------------------------
 
figure\,5:  Blend-by-blend (BB) analysis of GJ412A (p.17).

figure\,6:  BB analysis of HIP\,57050 (p.18).

figure\,7:  BB analysis of GJ406 (p.18).

\vspace{2mm}

table\,8: The result of the BB analysis for 42 M dwarfs (p.29).

table\,14: EWs of CO blends (for electronic version) (p.33).

----------------------------------------------------------------------

\subsection{External Errors}

Besides the internal errors (probable errors) of the BB analysis 
mostly 0.05 - 0.10\,dex (table\,8), there are various sources of the
external errors. We here examine the errors due to the uncertainties of the
fundamental parameters.  For this purpose, we select three objects
representing early (GJ\,338A; dM0.5), middle (GJ\,436; dM3.5), and late 
(GJ\,406; dM6.5e) M dwarfs, and repeat the BB analysis
for the reference parameters, which we think to be close to the best
solution for each object, by the use of the appropriate model for the 
parameters (first line for each star in table\,9). The resulting
logarithmic abundance correction
$\Delta\,{\rm log}\,A_{\rm C}$ to the assumed initial value (given in
the parenthesis in the first column) is given in the 5-th column. Then, we
change $T_{\rm eff}$ by $\pm 50$\,K (second \& third lines for each object), 
log\,$g$ by $\pm 0.25$ (4- \& 5-th lines), and
$\xi_{\rm micro}$ by $\pm 0.5$\,km\,sec$^{-1}$ (6- \& 7-th lines)
from those of the reference model, and carry out the BB analysis
by the use of the appropriate model for the changed parameter. The resulting  
logarithmic abundance correction  and the difference from that for the
reference model are given in the 5-th and 6-th columns
respectively, in table\,9 for each object. We also 
examine the effect of changing the Cc-series, which we are using throughout
the seven cases so far, to the Ca-series with the fundamental parameters
of the reference model in the first line for each object (8-th line for 
each object in table\,9).

Inspection of table\,9 reveals that  uncertainties of $\pm 50$\,K in
$T_{\rm eff}$ result in rather modest effects smaller than (in early M) or
comparable with (late M dwarf) the internal errors. This result can be
understood if we remember that CO formation is already complete (i.e., almost
all the carbon is in CO) and hence its abundance remains
almost unchanged for the changes of temperatures. On the other hand,
uncertainties of $\pm 0.25$\,dex in log\,$g$ result in an appreciable effect
especially in earlier M dwarfs. This may reflect the pressure dependence
of the background opacities due to the $f-f$ transitions of H$^{-}$ and
H$_2^{-}$ and collision-induced dipole transitions of H$_2$, which are
larger at  higher gravity and hence require a positive abundance correction
(i.e. larger carbon abundance).   
In the late M dwarf GJ\,406, CO lines are pretty strong (see figure\,13)
and larger pressure broadening under the higher gravity may require
negative abundance correction (i.e. smaller carbon abundance is sufficient
to account for the observed EWs).  The uncertainties of 
$\pm 0.5$\,km\,sec$^{-1}$ in the micro-turbulent velocity
 $\xi_{\rm micro}$ result in considerable changes in the carbon
abundances. This result implies that the problem of the micro-turbulent 
velocity remains to be important  even in the spectral analysis of M
dwarfs in which the pressure broadening  plays significant role.  
We could not determine the micro-turbulent velocity in this study, 
and this is the major shortcoming that we hope to conquer
in the near future. Finally, larger C \& O abundances in the Ca-series
compared to the Cc-series result in higher temperatures in the photosphere 
due to the increased line blanketing effect, and hence have a similar 
effect as the increased $T_{\rm eff}$ (compare the third and 8-th lines 
in each object in table \,9).

Another source of external error is the overall metallicity
which we assumed to be the solar for all the models. 
As an example of low metallicity case, we examine GJ412A which shows
a large decrease of carbon ($\Delta\,{\rm log}\,A_{\rm C}$ = -0.181; 
table\,8). For this purpose, we generate a model in which all the metal 
abundances are reduced by 0.20\,dex from the solar and repeat the BB 
analysis with the resulting model. The result shown in the last line 
in table\,9 suggests that the uncertainty  due to the use of solar metallicity
model is 0.094\,dex. The effect of the reduced metallicity on the
thermal structure of the model is a decrease   of temperatures due
to the less efficient backwarming effect and has a similar effect as the
case of the decreased $T_{\rm eff}$.

\vspace{2mm}
----------------------------------------------------------------------         

table\,9: Effect of the external errors on abundance determinations (p.30).

---------------------------------------------------------------------- 

\section{Analysis of the Spectra with the Depressed Continua by
the Molecular Veil Opacities}

   In the observed spectrum, we measure CO lines by referring 
to the pseudo-continuum, but 
analyzed the observed equivalent width measured in this way 
by the use of the predicted one evaluated by referring to the true
continuum (subsection 4.2). This analysis is obviously not self-consistent. 
As a possibility to resolve such inconsistency,
we analyze the spectra by referring to the pseudo-continua both in
observed and predicted spectra (subsection\,5.1).
We then consider the effect of the contamination of weak H$_2$O lines in 
the computation of the predicted EWs, with which we  analyze the 
observed EWs including the blending of weak H$_2$O lines as well. In this 
way, the consistency of the analysis can be recovered (subsection\,5.2).
Finally, we examine the
results by the spectral synthesis and $\chi^{2}$-test (subsection\,5.3).

\subsection{Effect of  the H$_2$O Veil Opacity on the Continua}

To examine the results of the BB analysis outlined in section\,4,
the spectral synthesis (SS) method can be of some use. 
For this purpose, we first show a theoretical spectrum of the CO (2-0)
bandhead region  for our model Cc2800c517 
 applied to GJ\,406 in figure\,8a. The spectrum is first
calculated at the sampling interval of 0.02\,\AA ~ or $R \approx
10^{+5}$ (thin line) and then convolved with the slit function
of the spectrograph (Gaussian) with FWHM = 16\,km sec$^{-1}$ (thick
line). Hereafter, we always show this low resolution version of the
theoretical spectrum alone
for simplicity throughout this paper. In the predicted spectrum,
the true continuum level is obviously known but cannot be seen at all
in figure\,8a,  and we know from figure\,4b
that this is entirely due to numerous weak lines of H$_2$O. 
However, the upper envelope of the spectrum
can be well defined as shown by the dashed line, as we already know in
the case of H$_2$O lines alone in figure\,4b. This means 
that the pseudo-continuum level can be well defined even though
the  ``continuum'' level is depressed by about 8\% for the low
resolution spectrum. 
 
In the observed spectra of M dwarfs, it is only possible to draw a 
pseudo-continuum. Given that it is possible to define the pseudo-continuum
accurately in the  predicted spectrum as in figure\,8a, 
we can compare the observed and predicted spectra both normalized by
their pseudo-continua.
Thus we renormalize the predicted spectrum in figure\,8a by its pseudo-
continuum (dashed line in figure\,8a)  and the result is shown in figure\,8b 
(thick line) in comparison with the observed spectrum also normalized
by its pseudo-continuum (filled circles)
\footnote{Hopefully the pseudo-continua for the observed and predicted 
spectra can be drawn consistently. Practically, pseudo-continuum 
for the predicted spectrum for a short interval can be a straight line
passing through the highest peak and hence rather simple. On the other
hand, the case of the observed spectrum may not be so simple since the 
pseudo-continuum level may suffer the effect of the variations due to 
the atmospheric transmission, atmospheric absorption, detector sensitivity etc.
Of course, these effects are corrected for during the data reductions,
but some effects may remain uncorrected.
Thus, the problem of the reference ``continuum'' is still a troublesome 
problem even  by the use of the pseudo-continuum.}. 
The observed (filled circles) and predicted (thick line) spectra both 
normalized by their pseudo-continua show a reasonable match, but 
the observed spectrum cannot be matched at all with the predicted
spectrum normalized by its true continuum ( simply
copied from figure\,8a above but shown by thin line).
 Hereafter, we always compare the observed and predicted spectra,
both normalized by their pseudo-continua. 

In the comparison outlined above, we assume log\,$A_{\rm C}$(BB) = -3.55
(table\,8) for GJ\,406 and log\,$A_{\rm O} = {\rm log}\,A_{\rm C} + 0.30$. 
The effect of oxygen abundance on CO spectrum may not be important
since CO abundance is almost determined by the carbon abundance alone.
However, if carbon abundance turns to  be very large in some M dwarfs and if
oxygen abundance is kept at its initial solar value, then  it may happen that 
 $A_{\rm C} > A_{\rm O}$ or the M dwarf turns to the dwarf carbon star! 
To prevent such a catastrophe,  we always adjust the oxygen abundance
to be $A_{\rm O}/A_{\rm C} = 2.0 $ for the given carbon abundance
throughout this paper. It is to be noted that this O/C ratio is
assumed to be  the same as the solar value (e.g.. \cite{All02}).

Now, as examples of comparing the observed and predicted
spectra this way, we show the cases of GJ\,412A and HIP\,57050
in figures\,9a and 9b, respectively. The thick lines are predicted 
spectra for the final carbon abundances  of
our BB analysis, ${\rm log}\,A_{\rm C}^{(2)}$, and the thin 
lines for ${\rm log}\,A_{\rm C}^{(2)} \pm \delta (=0.3)$. The observed spectra
normalized by their pseudo-continua (shown by the filled circles)  appear to 
agree rather well with those predicted for ${\rm log}\,A_{\rm C}^{(2)}$ 
(thick lines). We also evaluate $\chi^{2}$ values by
$$ \chi^{2} = {1\over{N-1}}\sum_{i=1}^{N} 
{\Bigl(}{ {f_{\rm obs}^{i} - f_{\rm cal}^{i} }\over\sigma_{i}
}{\Bigr)}^{2},   \eqno(4) $$
where $f_{\rm obs}^{i}$ and  $f_{\rm cal}^{i}$ are observed and predicted 
spectra normalized by their pseudo-continua, respectively. 
$N$ is the number of data points and $\sigma_{i}$
is  the noise level estimated from the $S/N$ ratio in table\,1 
(assumed to be independent of $i$). Resulting $\chi^{2}$ values for GJ\,412A
are 7.719, 4.241, and 17.469 for $\delta$ = -0.3, 0.0,  and +0.3, respectively,
where $\delta$ is a modification applied to ${\rm log}\,A_{\rm C}^{(2)}$. 
For HIP\,57050, $\chi^{2}$ values are 8.070, 3.808, and 11.520  for     
$\delta$ = -0.3, 0.0, and +0.3, respectively. The $\chi^{2}$ values for 
${\rm log}\,A_{\rm C}^{(2)}$ of 42 M dwarfs are given in the 10-th
column of table\,8. In evaluating $\chi^{2}$ values, some spectral
regions dominated by absorption other than CO  are masked (shown by the
filled areas in figures\,8, 9, and 13).  
 
We must remember that our logarithmic carbon abundances 
${\rm log}\,A_{\rm C}^{(2)}$  themselves are determined without considering 
the effect of the H$_2$O  contamination (subsection4.2). Nevertheless,
the fittings of the observed and  predicted spectra, both normalized
by the pseudo-continua defined by the numerous weak  H$_2$O lines,
appear to be not so bad, and this may be simply because the depressions of
the continua by the H$_2$O blends are not very large
\footnote{The depressions of the continua of the predicted spectra are 
about 1, 3, 6, and 8 \% 
for the models of $T_{\rm eff} =$ 3800, 3500 (see figure\,4b), 3200, 
 and 2800 (see figure\,8a)\,K, respectively, but depend also on log\,$g$
(for example, about 10 \% for slightly higher gravity in $T_{\rm eff} =$
 2800\,K model). The amount of the depression can be known only for the
theoretical spectrum and never be known for the
observed spectrum, since the true-continuum level cannot be known.}.  
But we should consider the effect of the H$_2$O contamination in the 
determination of the carbon abundance itself and we discuss this problem 
next  in subsection\,5.2.

\vspace{2mm}
---------------------------------------------------------------------    
             
figure\,8a,b: Observed and predicted spectra of GJ\,406 (p.19).

figure\,9a,b: Observed and predicted spectra of  GJ\,412A and HIP\,57050 
(p.19).

--------------------------------------------------------------------- 
             
\newpage

\subsection{Blend-by-Blend Analysis Based on the Equivalent-Widths Measured on
the Synthetic Spectra}
 One method to take the effect of H$_2$O contamination in determination of
the carbon abundances based on CO analysis may be to apply the SS method, 
which is now possible to apply to the spectra
 whose true-continua cannot be seen, by the way outlined in the preceding 
subsection.  We have already computed the spectra for three values
of carbon abundances in subsection\,5.1 and applied the $\chi^2$-test for the
fitting of the predicted spectra to the observed spectrum. Just by
computations of  additional several spectra with their $\chi^2$ values,
carbon abundance can be obtained by  minimization of the $\chi^2$ values.
However, the SS method seems to be a bit too intricate  especially 
for our medium resolution spectra. In fact, there is little merit to apply 
such a method to medium resolution spectra blurred by the slit function  of 
FWHM as large as 16 km sec$^{-1}$, which washes out all the details of the 
spectra characterized by the velocity parameters (micro- and macro-turbulence, 
rotation etc.) generally smaller than 1\,km\,sec$^{-1}$. 
Moreover, the SS method is by no means free from its own limitations 
as will be discussed later (subsection\,6.1).
For these reasons, we  propose  a more simple method based on the equivalent 
width  measurements instead. 

  For this purpose, we proceed as in our preliminary analysis (subsection\,4.2)
but  the predicted equivalent widths (EWs) are calculated by including 
the effect of H$_2$O contamination and by referring to the pseudo-continuum  
instead of the true-continuum. This can be done in principle by using
the usual code of evaluating the EWs, but it is again a bit too intricate
to include thousands of weak H$_2$O lines in computing an EW of
a CO blend, by referring to the pseudo-continuum which is difficult to
know locally instead of the true-continuum which is obvious in 
the theoretical spectrum.
Instead we can proceed more easily by measuring the EWs on the synthetic
spectrum in the same way as we measure the EWs on the observed spectrum.
In this way, the effect of the H$_2$O contamination can be taken into
account automatically and the pseudo-continua can also be defined
consistently for the predicted and observed spectra. 

Actually, we calculate the synthetic spectra for log\,$A_{\rm C}^{(2)}$ in 
9-th column of table\,8 and
log\,$A_{\rm C}^{(2)} \pm \delta $ (some examples for the case of 
$\delta =0.3$ is shown in figures\,8a and 8b), and we measure the predicted
EWs, $W(\delta)$, for the CO blends listed in table\,7.   We now
include the lines of Ref. no. 13 and 14, since the effect of the H$_2$O 
blending can be corrected for, but we  exclude the line of
 Ref. no.\,1 which appears to be too close to the bandhead.  
Then we have mini curves-of-growth defined by $W(\delta)/\lambda$
vs. $\delta$ (e.g. $\delta$ =-0.3, 0.0, and +0.3), by which the  
logarithmic abundance corrections 
to log\,$A_{\rm C}^{(2)}$ required to explain the observed EWs can 
be obtained. We assume $\delta = \pm0.3$ at the beginning. However,
we find that the logarithmic abundance corrections do not exceed -0.1 in 
general. Then, we assume $\delta = -0.1$ and a linear interpolation
is sufficient to find the abundance correction for each
observed EW.   Thus our final
blend-by-blend analysis based on the EWs measured on the synthetic spectra
 (to be referred to as BBSS or simply as BS method) is quite simple.
 
As an example of the BS analysis, the resulting logarithmic abundance 
corrections  for the case of GJ\,412A 
are shown in figure\,10  against the observed EWs used, and we obtain
the mean value of $\Delta\,{\rm log}\,A_{\rm C}^{(3)} = -0.09 \pm 0.04 $
shown by the dashed line. This is the
correction to the carbon abundance obtained from CO blends
disregarding the effect of H$_2$O contamination in table\,8, and our iterative
abundance analysis now finishes  by this self-consistent
analysis taking the effect of H$_2$O blending into account. The resulting
final logarithmic carbon abundance is:
$$ {\rm log}\,A_{\rm C}^{(3)} =  {\rm log}\,A_{\rm C}^{(2)} +  
 \Delta{\rm log}\,A_{\rm C}^{(3)}.     \eqno(5) $$ 
Additional examples are shown in figures\,11 and 12  for HIP\,57050 and 
GJ\,406, respectively. We carried out this  analysis for our  42 M dwarfs 
and the resulting logarithmic abundance corrections and the final
logarithmic carbon abundances with the probable errors are given in the 
second  and third columns of table\,10, respectively. It is to be noted 
that the values of $\Delta\,{\rm log}\,A_{\rm C}^{(3)}$
are always negative for 42 M dwarfs, and this is because the carbon abundances 
overestimated by the BB analysis neglecting the effect of
contamination of H$_2$O lines are now corrected for by the BS analysis
taking this effect into account. 

Inspection of figures\,5 $\sim$ 7 reveals that the abundance corrections
by the BB analysis always increase with the observed EWs. 
Normally, the variations of the abundance corrections plotted
against the EWs used provide important information on the
physical parameters relating to the  line formation such as
the micro-turbulent velocity (e.g., \cite{Tsu08}). In the present
case, however, the lines used are confined to a restricted
intensity range  and the variations may simply be
explained as the effect of contamination mainly of H$_2$O lines.
In fact, EW will increase by the blend of H$_2$O lines and
a larger EW naturally results in a larger abundance correction,
exactly as shown in figures\,5 $\sim$ 7. This systematic effect  disappears in
figures\,10 $\sim$ 12 where  the  effect of H$_2$O contamination is properly 
considered.
Also, the abundance corrections, after the effect of H$_2$O contamination
is corrected, approach the lowest abundance
corrections in figures\,5 $\sim$ 7 obtained by disregarding the H$_2$O 
blending. This may be because such lines giving the lowest abundance
corrections in figures\,5 $\sim$ 7 are simply those  suffer least effect 
of H$_2$O  contamination.

\vspace{2mm}
--------------------------------------------------------------------- 
                
figure\,10: Blend-by-blend analysis by EWs measured on synthetic 
spectrum (BS method) of GJ\,412A (p.20).

figure\,11: BS analysis of HIP\,57050 (p.20).

figure\,12: BS analysis of GJ\,406 (p.20).

\vspace{2mm}

table\,10: The result of BS analysis for 42 M dwarfs (p.31).

--------------------------------------------------------------------- 

\subsection{Final Check by the Spectral Synthesis and $\chi^{2}$-Test }
   
    We generate the synthetic spectra for our final logarithmic carbon 
abundances $ {\rm log}\,A_{\rm C}^{(3)}$ in table\,10  and assuming the 
logarithmic oxygen abundances
of $ {\rm log}\,A_{\rm O} =  {\rm log}\,A_{\rm C}^{(3)} + 0.30$. 
We compare the observed (filled circles) and predicted (solid line) spectra 
for six  M dwarfs in figure\,13 and the fits are generally fine if not 
perfect. Inspection of figure\,13 reveals that
the change of the observed spectra from dM0 to dM6.5 can be well accounted for 
by the predicted spectra with $T_{\rm eff} \approx 2800 \sim 3900$\,K 
and $ {\rm log}\,A_{\rm C} \approx -3.8 \sim -3.2$.
By the way, a  feature at $\lambda \approx 22970$\,{\AA} 
was identified as due to Ti I a\,$^{3}$G$_{5}$ - z\,$^{3}$F$_{4}$ and
a blending feature at $\lambda \approx 22973$\,{\AA}  
to Sc I a\,$^{4}$F$_{9/2}$ - z\,$^{4}$F$^{0}_{9/2}$ on the
sunspot umbral spectrum \citep{Wal92}.  
The H$_2$O feature just shortward of 
the CO (2, 0) bandhead strengthens towards later M dwarfs as expected.

For GJ\,412A and HIP\,57050, we can compare figures\,13c and 13d with 
figures\,9a and
9b, respectively, showing the similar comparisons but with the results of
the BB analysis. The differences are rather minor, but the improvements
can be confirmed  by their $\chi^{2}$ values (see below). The $\chi^{2}$ 
values for the final abundances ${\rm log}\,A_{\rm C}^{(3)}$ by the BS 
analysis  are given in 4-th column of table.\,10 for the 42 M dwarfs, 
and the results show definite  improvements compared with those for 
the abundances ${\rm log}\,A_{\rm C}^{(2)}$ by the BB analysis
given in the 9-th column of table\,8 in general.
  
Finally, we evaluate the $\chi^{2}$ values for the synthetic spectra
assuming several carbon abundances and plot the results against the assumed 
logarithmic carbon abundances in figure\,14 for three M dwarfs, GJ\,412A, 
GJ\,406, and HIP\,57050. The points marked by $\pm n \delta$ ($\delta = 0.1;
 n = 1 - 3$) are $\chi^{2}$ values for the synthetic spectrum assuming 
$ {\rm log}\,A_{\rm C} =  {\rm log}\,A_{\rm C}^{(2)} \pm n \delta$ with 
log$\,A_{\rm C}^{(2)}$ for each object given in the 9-th column of table\,8.
The points marked by BB are $\chi^{2}$  values for our final carbon
abundances of the BB analysis while those by BS  for our BS analysis.
Inspection of figure\,14 reveals that our BS analysis results
in the $\chi^{2}$  value near the minimum  for each object.
Thus our simple BS analysis results in almost the same carbon abundance as by
the minimization of $\chi^{2}$ values.

\vspace{2mm}
---------------------------------------------------------------------  
                            
figure\,13:  Predicted vs. observed spectra for 6 M dwarfs (p.21).

figure\,14:  $\chi^{2}$-test  for GJ412A, GJ406, and HIP\,57050 (p.22).

---------------------------------------------------------------------
             
\section{Discussion}

\subsection{Method of Analysis}
The major obstacle in the abundance analysis of cool stars has been
the difficulty to locate the continuum level, and this problem is not limited
to M dwarfs but has been a serious problem in cool luminous stars including
M  giant and supergiant stars. We hope to relax this difficulty by 
analyzing the spectra by referring to the pseudo-continuum instead of the 
true-continuum. Certainly, the pseudo-continuum defined by the
collective effect of numerous weak lines is much more complicated
compared with the true-continuum defined by the free-free or bound-free
transitions of a few atoms and ions. However, 
the spectroscopic  analysis can be carried out  essentially the same way
by referring to the pseudo-continuum instead of the true-continuum.
For example, we measure EWs of the blended features by referring to 
the pseudo-continuum both in the observed and predicted spectra. 
Then, the analysis of EWs can be carried out as usual and this is
important since the analysis of EWs plays a significant role in the 
spectral analysis for the reason to be noted below. 

In the  analysis of the complicated spectra composed of many blended lines,
the spectral synthesis (SS) 
method has been widely employed, and the best fit has been
determined by the $\chi^{2}$-test in general. However, an
important drawback in this approach is that the different line
broadenings could not be separated well. Especially,
the micro-turbulent velocity could not be determined well since
its effect could not be separated from those of macro-turbulence
and rotation on the synthetic spectrum. For example, the SS method
was applied to several M dwarfs and  micro- and macro-turbulent
velocities were determined by \citet{Bea06} , but it is not very 
clear how these two velocities could have been separated.
Also, many groups used the SS method in a
comparative study of the spectra of cool giant stars but  
the turbulent velocities were not determined well in many
cases by the SS method \citep{Leb12}.

The important discovery of the micro-turbulent velocity 
was done from the analysis of the equivalent widths by the curve-of-growth
method \citep{Str34}, and it is the EWs that are directly 
influenced by the micro-turbulent velocity. On the other hand,
other line broadenings such as due to the macro-turbulence and rotation
give no effect on EWs and they are recognized only on
the synthetic spectrum. For this reason,  the micro-turbulent velocity
can be determined accurately and easily from the analysis of EWs, as
has been shown for the case of cool giant stars (e.g., \cite{Tsu08}; 
\yearcite{Tsu09}). Once the micro-turbulent velocity is determined, 
then the SS method can be used to infer the effect of the  additional 
broadenings such as the macro-turbulence and rotation. Certainly,
SS method is not useful as a means by which to determine all the  parameters 
but it can be useful to check the overall consistency at the end.    

At the beginning of this study,
we hoped to determine the micro-turbulent velocity first from
the analysis of EWs by the LL method, but this attempt could not
be realized with the medium resolution spectra at hand  (subsection\,4.1). 
Even with the higher resolution, the effect of the blending makes the LL 
method difficult. For this purpose, we propose to analyze the EWs measured 
on the synthetic spectra in subsection\,5.2 and this method, referred to as 
BS method, works well in determining the abundance. By this method,
analysis of EWs is not necessarily limited to  single lines but can be
extended to blended lines found in the region of depressed continua. 
We hope to apply  this method to determining the micro-turbulent velocity
with the higher resolution spectra.  This is certainly a more 
challenging problem, but this should be almost unique way to determine
the micro-turbulent velocity in the M dwarf photospheres.    
 
\subsection{Carbon Abundances in M Dwarfs}
  Determination of carbon abundances in late-type stars has been
difficult in general and we do not yet have a result generally accepted
even for the Sun. This is because the spectral lines used as indicators
of carbon abundance are highly sensitive not only to the thermal
structure of the photosphere but also to the inhomogeneity and non-LTE 
effects etc.  In contrast, such difficulties can be avoided  in M dwarfs
for which CO molecule can be used as the abundance indicator. In the
cool and dense photospheres of M dwarfs, almost all the carbon is
in CO and hence its abundance shows little change for the 
changes of the physical condition. This advantage
cannot be applied to hotter stars such as the Sun, in which CO formation is 
not yet complete but changes its abundance drastically for a minor
change of temperature\footnote{In sufficiently low temperatures,
say $T \lesssim 4000$\,K at log\,$P_{\rm g} \approx 6.0$, CO formation
is complete and the partial pressure of CO, $P_{\rm CO} \approx 
A_{\rm C}P_{\rm g}$, is almost independent of $T$. In the region where 
CO formation is incomplete (i.e., $P_{\rm CO} << A_{\rm C}P_{\rm g}$),
the partial pressures of free C and O are $P_{\rm C} \approx
 A_{\rm C}P_{\rm g}$ and $P_{\rm O} \approx A_{\rm O}P_{\rm g}$, 
respectively. Then,
$P_{\rm CO} = P_{\rm C}P_{\rm O}/K_{\rm CO}(T) \propto A_{\rm C}A_{\rm O}
P_{\rm g}^{2} 10^{ {5040 \over T}D_{\rm CO} }$ where $K_{\rm CO}(T)$ is the
equilibrium constant defined in \citet{Her45}  and, with the large
dissociation energy of CO, $ D_{\rm CO} = 11.09$\,eV, $P_{\rm CO}$
shows a drastic change with $T$.}.  For this reason, CO cannot be
a good abundance indicator of carbon for the Sun \citep{Tsu77}. 
Despite such a difficulty, however, an extensive analysis on the solar model 
photoshpere with the use of CO spectrum itself  has been done and 
 the solar carbon, oxygen, and their isotopic abundances have been
 determined from the CO ro-vibrational spectrum \citep{Ayr06}.

Given that the difficulty due to the depressed continuum could be
overcome by referring to the pseudo-continuum, the advantage
of CO as the abundance indicator of carbon for M dwarfs outlined above
can fully be realized.
In fact, we have determined the carbon abundances from CO in 42 M dwarfs
rather easily after correcting for the effect of the contamination of
numerous weak H$_2$O lines (section 5). So far, determinations of
the elemental abundances in M dwarfs by a direct spectroscopic analysis are
rather scarce, and our result for carbon abundances in 42 M dwarfs may be 
the largest sample of a directly determined elemental abundance or
metallicity in M dwarfs at present. 

An unsettled  problem is that we have
two cases of different $T_{\rm eff}$ values for GJ\,273 and GJ\,725B. 
As for GJ\,725B, the resulting carbon abundance  of GJ\,725B-H 
(log\,$A_{\rm C}^{(3)} = -3.61 \pm 0.08$)  agrees rather well with that of 
GJ\,725A (log\,$A_{\rm C}^{(3)} = -3.58 \pm 0.09$) while that of GJ\,725B-L 
(log\,$A_{\rm C}^{(3)} = -3.86 \pm 0.08$) does not (see table\,10).
Although such a test by binary cannot be applied to GJ\,273, the result
of the binary test for GJ\,725B suggests that the $T_{\rm eff}$
by the $M_{\rm 3.4}$-method should be preferable for GJ\,725B and hence 
possibly for GJ\,273. Also, the deviations 
from the mean relations in figures 1 $\sim$ 3 are rather large for
GJ\,273-L as well as for GJ\,725B-L. For these reasons, we adopt the
results for GJ\,273-H and GJ\,725B-H in the following discussion. 

We compare our carbon abundances with the metallicities determined from
the infrared spectroscopy by \citet{Mou78} and by \citet{One12}
in table\,11 . The error bars were not given explicitly by \citet{Mou78} 
and we estimate them to be $\pm$ 0.2\,dex from what is mentioned in his
text. We also include two representative solar carbon abundances so far 
proposed, but the values somewhat between these extreme values
have also been suggested (e.g., \cite{Ayr06}; \cite{Asp09}). We plot these
data in figure\,15a, and our log\,$A_{\rm C}$ values show expected positive 
correlation with [M/H] as a whole.
Close inspection of figure\,15a reveals that there is  bifurcation in the
log\,$A_{\rm C}$ - [M/H] relation: one branch includes a larger sample of
M dwarfs and the Sun with the high carbon abundance \citep{Gre91} while
the other a smaller sample of M dwarfs and the Sun with the low carbon 
abundance \citep{All02}. 

In table\,12,  we compare our carbon abundances with the iron abundances
 determined from the photometric calibrations of [Fe/H] using  high
resolution spectra \citep{Nev13}, and plot these data together with 
the solar values from table\,11 in figure\,15b.  Except for two deviating 
stars (GJ\,406 and GJ\,686),
the  log\,$A_{\rm C}$ - [Fe/H] correlation suggests that the C/Fe ratios in
M dwarfs are nearly constant at about the solar C/Fe ratio based
on the high carbon abundance \citep{Gre91} rather than on the
downward revised carbon abundance \citep{All02}. This conclusion applies
to the majority of M dwarfs shown in figure\,15a as well. Given that the
determination of carbon abundances is rather difficult in late type stars
as exemplified by the solar case and that it is rather easy in 
M dwarfs by the use of CO as noted above, our result on the
carbon abundances in M dwarfs provides a strong constraint on the carbon
abundances of the disk population in the solar neighborhood. 

Presently known data shown in figure\,15 suggest a possibility of 
bifurcation in C/Fe ratio in M dwarfs. The majority of M dwarfs belong
to the  group of C/Fe ratio with the high solar carbon abundance and the
minority belong to another group of C/Fe ratio with the low solar
carbon abundance. It may probably not be  a serious problem if few
M dwarfs belong to the minority group. However, it should be a
problem if the Sun belongs to the minority group or the solar C/Fe
ratio is atypical for [Fe/H]=0. This problem if the Sun is a typical
star with respect to the relative abundances of the elements at [Fe/H]=0
was examined in detail by the analysis of 189 nearby unevolved stars
by \citet{Edv93}, who concluded that the Sun is a quite typical star
with respect to the relative abundances of about a dozen of the elements
for its metallicity, age, and galactic orbit, although carbon itself was not  
included in their analysis. We point out that  the recent downward revised 
solar carbon abundance is apparently contradicting with this result.  

Finally, the distribution of
42 M dwarfs  against the carbon abundances is shown in figure\,16, and
most M dwarfs are confined to the range of log\,$A_{\rm C}$ between -3.7 and
-3.2 except for few cases outside of this range. Also, our sample   
includes ten planet hosting M dwarfs (those marked by $\ddagger$ in
table\,1) as illustrated in figure\,16. 
It appears that the planet hosting M dwarfs are biased towards
carbon rich side while M dwarfs as a whole towards carbon
poor side.   As a result, the fraction of the planet hosting M dwarfs is 
larger in M dwarfs of higher carbon abundances for our present sample. 
This result is in agreement with a tendency of metallicity in G, F, K, and M
stars (e.g., \cite{Fis05}; \cite{Joh09}). 
However, our sample  cannot be regarded as an unbiased sample yet
and  we  defer a detailed discussion on this matter
to a future paper hopefully on an extended sample of M dwarfs. 

\vspace{2mm}
--------------------------------------------------------------------- 

figure\,15a,b:  Carbon abundances and metallicities in M dwarfs (p.22).

figure\,16: Distribution of carbon abundances in 42 M dwarfs (p.23).

\vspace{2mm}

table\,11: Carbon abundances and metallicities in M dwarfs (p.32). 

table\,12: Carbon and iron abundances in M dwarfs (p.32). 

--------------------------------------------------------------------- 

\subsection{Accuracy of the  Abundance Analysis}

The major problem in the spectral analysis of cool stars
is generally thought to be due to the difficulty to locate the continuum level.
We have examined the effect of depressed continuum due to the
H$_2$O veil opacity by our BS analysis and compared with the BB
analysis disregarding the effect of H$_2$O contamination. We have found that
the neglect of the H$_2$O blending results in an error of about
0.1\,dex at the largest in the derived abundance 
(see $\Delta\,{\rm log}\,A_{\rm C}^{(3)}$ in table\,10). 
The probable errors of the BB analysis are 0.05 - 0.1\,dex
(table\,8) and those of the BS analysis are slightly larger (table\,10). 
Thus, the effect of the depressed continuum may not be very large
in the case we have studied in this paper. 

We estimate the external errors due to the uncertainties in the fundamental
stellar parameters by the BB analysis (subsection\,4.3), and the results
may be the same for the BS analysis, since the effects of the
fundamental parameters should not be different for the two analyses.
The uncertainties in log\,$g$
and in $\xi_{\rm micro}$ result in rather large external errors in
general. However, we used the specified model for
the log\,$g$ of each M dwarf and we hope that the effect of 
uncertainty in log\,$g$ can be minimized. The problem is the
uncertainty in $\xi_{\rm micro}$: 
If the results of $\xi_{\rm micro} = 0.83 \sim 0.94$\,km\,sec$^{-1}$
\citep{Bea06} are typical values for M dwarfs, the uncertainty may not
be so large as assumed in subsection\,4.3. Anyhow,  
this is the largest problem in our present analysis and we hope to determine 
$\xi_{\rm micro}$ by  higher resolution spectra in future.  

We assume  simple one-dimensional LTE models
throughout this paper. Certainly, this is an over-simplification
to the real photospheres of M dwarfs. For example, convection
penetrates to the line forming region even in the classical
convective models based on the mixing-length theory (e.g., \cite{Tsu02}). 
However, even if inhomogeneity due to convection appears, the CO
abundance suffers little change since CO is well stabilized
in the photospheres of M dwarfs. For this reason, the inhomogeneity
will not have such a drastic effect as in the Sun (e.g., \cite{All02}), at 
least for the analysis of CO lines in M dwarfs. Also, line formation in the 
dense photospheres of M dwarfs can be treated within the framework of LTE, 
especially for the ro-vibration transitions \citep{Hin75}.

Finally, we first wondered if abundance analysis could be done with
the resolution as low as 20000. We have found, however, that the
abundance analysis is rather simple at such a medium resolution, as can be
known from the fact that the synthetic spectra can be matched rather
easily without specifying any velocity parameters to the observed spectra
in which all the details are smeared out by the slit function of
FWHM = 16\,km sec$^{-1}$. Such a fitting does not provide any
new information except for confirming the abundance determined
by the analysis of the EWs, but such a simplicity of analysis will be
useful if we are to analyze a large sample of stars.
We are convinced that the abundance analysis can be carried out by an 
analysis of EWs with the spectra of medium resolution, even though 
a high accuracy as realized by the higher resolution (e.g. 
$R \gtrsim 50000$) cannot be achieved. However,
even if the highest accuracy cannot be attained, the chemical composition 
is the prime fundamental data in astronomy and should best be  determined 
by the direct analysis of the spectra. We believe that the spectroscopic 
analysis of M dwarfs  can provide  unique  contribution to the problem of  
the cosmic  chemical abundances and we hope more efforts will be directed 
to this field.

\subsection{Hertzsprung-Russell Diagram at the End of the Main Sequence}
  The $M_{\rm 3.4} -{\rm log}\,T_{\rm eff}$ diagram shown in figure 1
reveals a characteristic feature in that it shows bendings at 
${\rm log}\,T_{\rm eff} \approx 3.56$ and $\approx 3.5$. 
Such  features should reflect the intrinsic properties of the stellar 
evolutionary models and, to clarify such a relationship,  we  generate an 
$M_{\rm bol} - {\rm log}\,T_{\rm  eff}$ 
diagram or  HR diagram by the use of the data given in tables 3 and 4.
The result is shown in figure\,17 where data based on the interferometry 
(filled circles) are extended to the lower temperatures by the data based 
on the infrared flux method (open circles). For comparison,  the
theoretical HR diagram based on the evolutionary models by \citet{Bar98} is 
shown by the solid line, on which stellar masses (in unit of $M_{\odot}$) 
of the models are indicated. 
Their models assumed the solar metallicity and covered the range between  
0.075 and 1.0 $M_{\odot}$, thus reaching the very end of the main sequence.
Although the number of data points is rather small, observations also 
cover the same range. The agreement between the theory and observations 
is fine especially in the regime of M dwarfs including the coolest ones.
   
The bendings noted on figure\,1 are clearly transformed from those in 
figure\,17.  The changes of the slopes in figures\,1 and 17 correspond
nicely to those in  the central temperature - effective temperature
relation given in Fig.\,1 of \citet{Bar98}. These changes of the slope
have been explained as due to the onset of convection induced by the 
H$_2$  formation in the photospheres of $\approx 0.5\,M_{\odot}$ 
dwarfs ($T_{\rm eff} \approx 3600$\,K)
and increased importance of the electron degeneracy in the interior of
$\approx 0.2\,M_{\odot}$ dwarfs ( $T_{\rm eff} \approx 3200$\,K)
\citep{Bar98}. These points roughly correspond to the 
bendings at  log\,$T_{\rm eff} \approx 3.56$ and at log\,$T_{\rm eff} 
\approx 3.5$ in figures\,17 as well as in figure\,1. Thus, recent 
observational HR diagram reaching the bottom of the main sequence
finally provides fine confirmation on the detailed structures of the 
sophisticated evolutionary models of low mass stars, which had been 
accomplished long time ago.   

\vspace{2mm}
---------------------------------------------------------------------  
                    
figure\,17: Observed and theoretical HR diagrams for low mass stars (p.23).

--------------------------------------------------------------------- 
           
\section{Concluding Remarks}
In this paper, we have shown that  the pseudo-continuum can be evaluated
accurately on the theoretical spectrum of the M dwarf and use it as a 
reference in comparison with the observed spectrum for which only the 
pseudo-continuum can be known. Then quantitative analysis of the spectra 
badly depressed by the molecular veil opacities can be carried out to 
some extent. Such an analysis could be made possible by the recent progress in 
molecular physics which provided extensive line-lists with 
high precision (e.g., \cite{Bar06}; \cite{Rot10}), and
the importance of the molecular databases should be emphasized again.

  Given that the difficulty due to depressed continuum  has been 
overcome,  abundance analysis of  M dwarf stars offers unique
opportunity which is not realized in other spectral types. In 
particular, we are now convinced that  at least the carbon
abundances in late-type stars could best be determined in M dwarfs
rather than in solar type stars by the use of CO as the abundance 
indicator. This result may be somewhat unexpected, but this is due to a 
favorable circumstance  that carbon atoms
in M dwarfs are mostly in stable CO molecules which remains
almost unchanged for the changes of physical condition 
in the photospheres of M dwarfs. 
The similar favorable condition is realized in the determination of
oxygen abundance by the analysis of H$_2$O consuming most oxygen 
left after CO formation and hence stable in the photospheres of 
M dwarfs. The accurate determination of carbon and oxygen abundances 
in M dwarf stars, representing those of the stellar components of the 
Galactic disk, has a special interest in connection with the yet unsettled 
problem of the solar carbon and oxygen abundances.  

  In this work, we mainly observed M dwarfs earlier than dM5
except for dM6.5 dwarf GJ\,406, since we thought that it might be difficult
to analyze later M dwarfs. However, we find that the dM6.5 dwarf
GJ\,406 could be analyzed in the same way as in the earlier M dwarfs
without any additional problem. Thus we are now convinced
that the late M dwarfs can also be analyzed similarly only if  
the pseudo-continuum level can be defined. In the coolest
M dwarfs of $T_{\rm eff} \lesssim 2600$\,K, a new problem is
that dust should form in their photospheres (e.g., \cite{Tsu96}: 
\cite{Jon97}). We are prepared to extend our analysis to
such a case with our Unified Cloudy Models (UCMs) accommodating
the dust clouds formed in their photospheres (\cite{Tsu02};
\yearcite{Tsu05}). We hope that such an  extension to the coolest M dwarfs
will be a pilot study for future works on more dusty objects
including the exoplanets themselves.  

  Finally, we regret that our analysis is not yet satisfactory in that
the spectral resolution ($R \approx 20000$ or velocity resolution of
16\,km\,sec$^{-1}$) is not high enough to extract the basic information  
coded in the spectra of M dwarfs (e.g., turbulent velocities) and we
hope further progress in  high resolution infrared spectroscopy in the 
near future. The scientific justification for such a quest
is obvious. In fact, there are large fields unexplored by high resolution
in low luminosity  objects including  M subdwarfs, dwarf carbon 
stars, L, T, and Y dwarfs, and exoplanets of various kinds, even if we
confine our attention to the objects somewhat related to M dwarfs. 
The high resolution spectroscopy on these objects will be within the 
capability of the large telescopes both on ground and in space under 
planning (or already in construction) only if efficient spectrographs 
can be developed.

\bigskip

We thank Y. Takeda and the staff of the Subaru Telescope for their
help in observations. We also thank Y. Takeda for his help in data
reduction of the echelle spectra. 

We thank an anonymous referee for careful reading of the text and for 
helpful suggestions, especially for suggesting to comment on the 
reason for the bendings in the plot of figure\,1.

This research makes use of data products from the Wide-field Infrared
Survey Explorer which is a joint project of the University of
California, Los Angeles, and the Jet Propulsion Laboratory / 
California Institute of Technology, funded by NASA.

This research has made use of the VizieR catalogue access tool and
the SIMBAD database, both operated at CDS, Strasbourg, France, and
of the RECONS database in www.recons.org.

Computations are carried out on common use data analysis 
computer system at the Astronomy Data Center, ADC, of the National 
Astronomical Observatory of Japan.

\newpage

\appendix

\section{On the New Effective Temperatures by Interferometry}
During the preparation of this manuscript, we notice a paper by \citet{Bra14} 
who reported new measurements of stellar diameters by interferometry,
and we find that the effective temperatures of three M dwarfs included in
our sample have been determined anew by their work. We compare their
results with ours based on the $M_{\rm 3.4}$ - log\,$T_{\rm eff}$ relation 
in table\,13.

We are a bit shocked to find a large discrepancy for GJ\,876, but we
are also worried to find that the $T_{\rm eff}$ of this object by the 
interferometry deviates from the mean relation in figure\,1 as the case of 
GJ\,725B.  In fact, the situation of GJ\,876 is very 
similar to the case of GJ\,725B for which interferometry gave $T_{\rm eff} 
= 3104$\,K while the $M_{\rm 3.4}$-method suggests $T_{\rm eff} = 3337$\,K,
and we finally decide to adopt  the higher value based on the 
$M_{\rm 3.4}$-method by considering the result of a binary test 
(see subsection\,6.2). Presently, we have no 
solution for GJ\,876 and we leave this problem open.

\vspace{2mm}
---------------------------------------------------------------------
                  
table\,13: New $T_{\rm eff}$'s by the interferometry (p.32).   

---------------------------------------------------------------------

\clearpage

\onecolumn

\begin{figure}
   \begin{center}
       \FigureFile(160mm,115mm){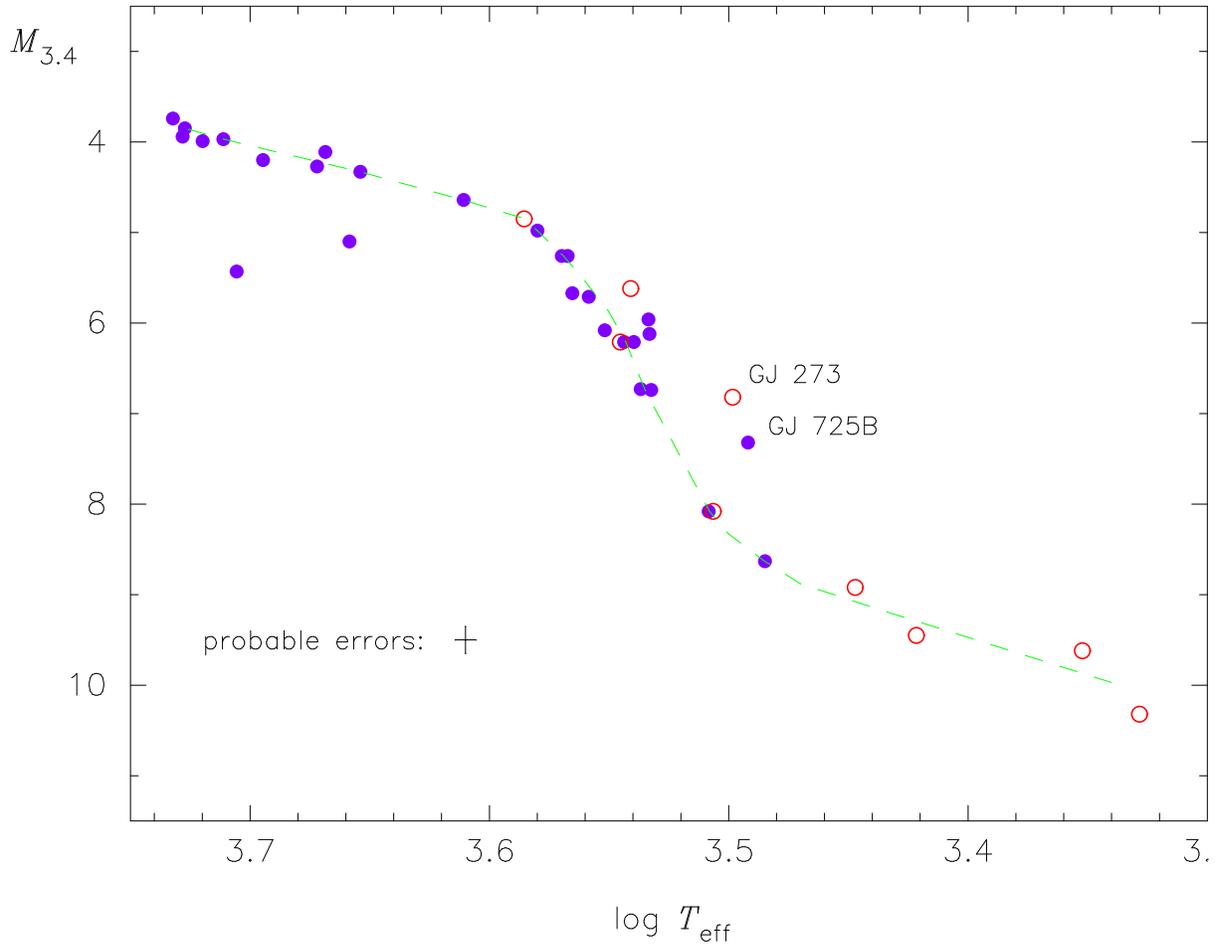}  
   \end{center}
   \caption{The absolute magnitude at 3.4\,$\mu$m, $M_{\rm 3.4}$, plotted
 against log\,$T_{\rm eff}$, where $T_{\rm eff}$ values are based on the
interferometry (filled circles) and infrared flux method (open circles).
The dashed line is a mean curve. 
The probable errors of $T_{\rm eff}$, reproduced from Table\,6 of \citet{Boy12}
to our table\,3, are less than 1\,\% for 21 objects out of 27 objects and 
only one object shows larger than 2\,\% error in the remainning 6 objects. 
The probable errors of $M_{\rm 3.4}$ are less than 0.15\,mag for 22 objects 
out of 27 objects (see table\,3). The error bars representing 1\,\% 
probable error in $T_{\rm eff}$ and probable error of 0.15\,mag in 
$M_{\rm 3.4}$ are shown at the lower left corner. 
}\label{figure1}
\end{figure}

\begin{figure}
   \begin{center}
       \FigureFile(80mm,70mm){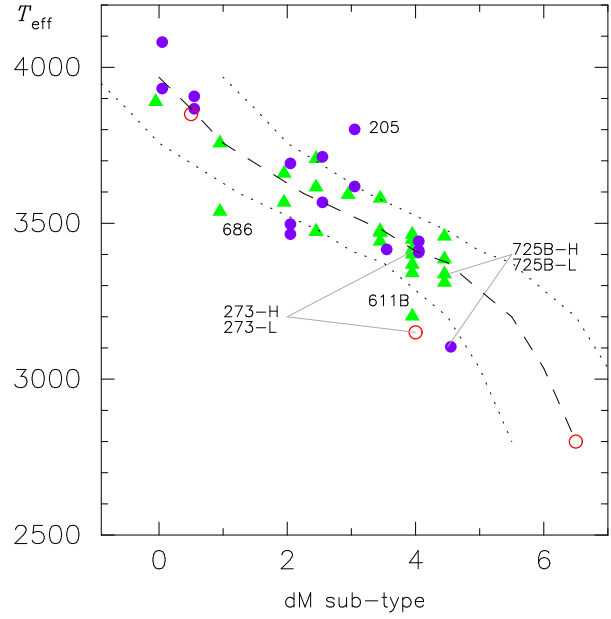}  
   \end{center}
   \caption{$T_{\rm eff}$ values  based on the interferometry (filled circles), 
infrared flux method (open circles), and $M_{\rm 3.4}$-method 
(filled triangles) are plotted against sp. types mostly (29 objects) by 
\citet{Joy74} and partly (13 objects) by SIMBAD. Most stars are found
 within the dispersion of $\pm$1 subtype (dotted lines) around the
mean curve (dashed line). The GJ numbers of objects showing large
deviations are indicated. 
}\label{figure2}
\end{figure}

\begin{figure}
   \begin{center}
       \FigureFile(80mm,70mm){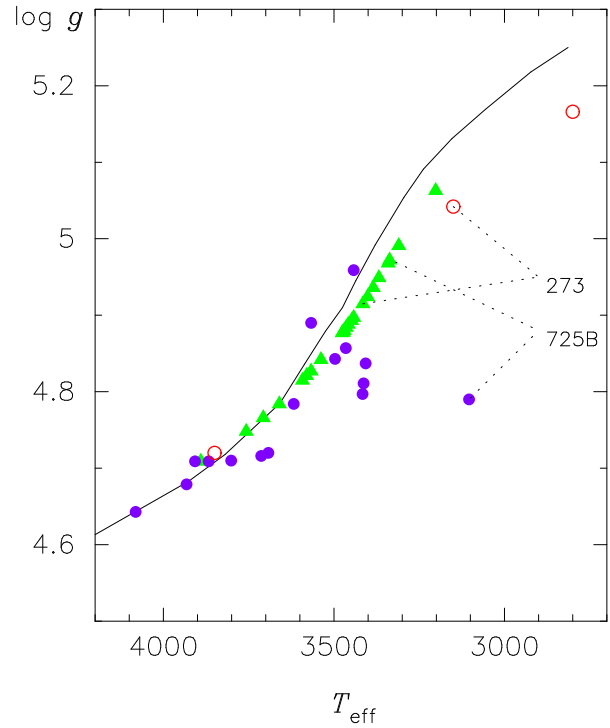}  
   \end{center}
   \caption{Log\,$g$ plotted against $T_{\rm eff}$, where $T_{\rm eff}$ values 
are based on the interferometry (filled circles), infrared flux method
 (open circles), and $M_{\rm 3.4}$-method (filled triangles). The solid line 
is based on the evolutionary models by \citet{Bar98}.}\label{figure3}
\end{figure}

\begin{figure}
   \begin{center}
       \FigureFile(145mm,90mm){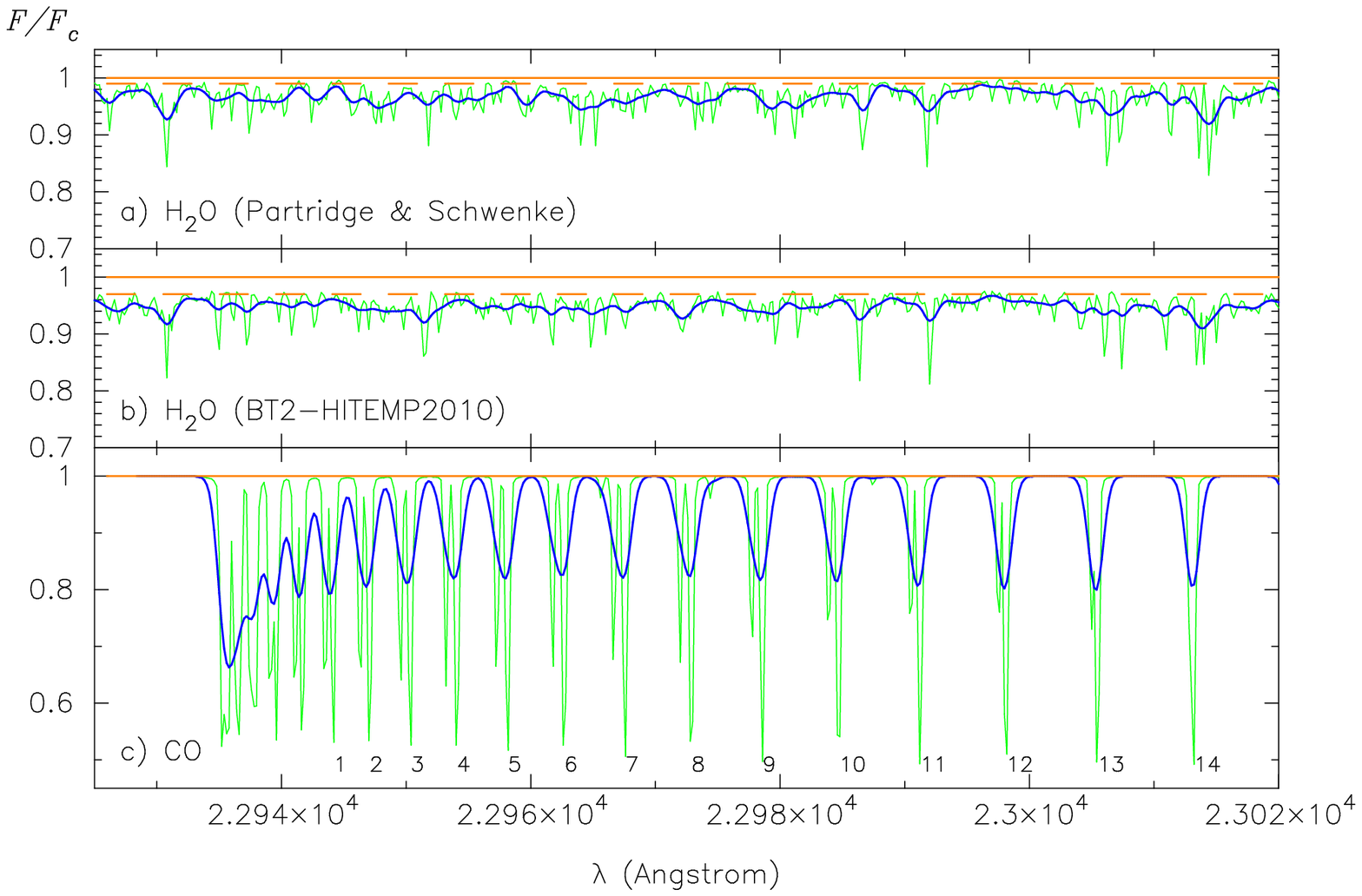}  
   \end{center}
   \caption{Theoretical spectra for the model Cc3500c475. The spectra
 are first calculated with a sampling interval of 0.02\,\AA (thin lines) 
and then they are convolved with the slit function of 
FWHM = 16\,km\,sec$^{-1}$ (thick lines). The true- and pseudo-continuous
levels are shown by the solid and dashed lines, respectively. 
a) H$_2$O spectrum based on the line-list by \citet{Par97}. b) H$_2$O 
spectrum based on the line-list BT2-HITEMP2010 (\cite{Bar06}; \cite{Rot10}). 
c) CO spectrum near the (2,0) band head. The reference numbers of
 table\,7 are given to respective CO blends.
  }
\label{figure4}
\end{figure}

\begin{figure}
   \begin{center}
       \FigureFile(80mm,75mm){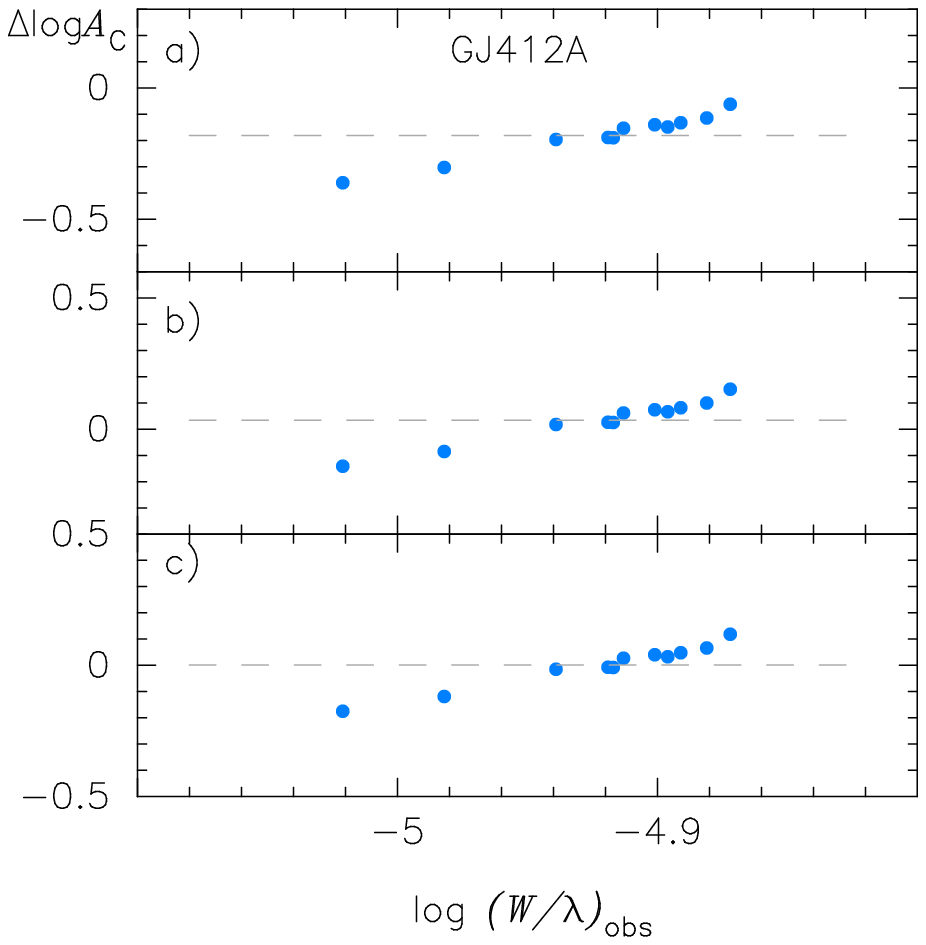}
   \end{center}
   \caption{The resulting logarithmic abundance corrections by the BB analysis 
for the CO lines in  GJ\,412A plotted against the observed values of
log\,${W/\lambda}$. The dashed line shows the mean correction. 
a) First iteration.  b) Second iteration. c) Confirmation of the convergence.
    }
\label{figure5}
\end{figure}

\begin{figure}
   \begin{center}
       \FigureFile(80mm,75mm){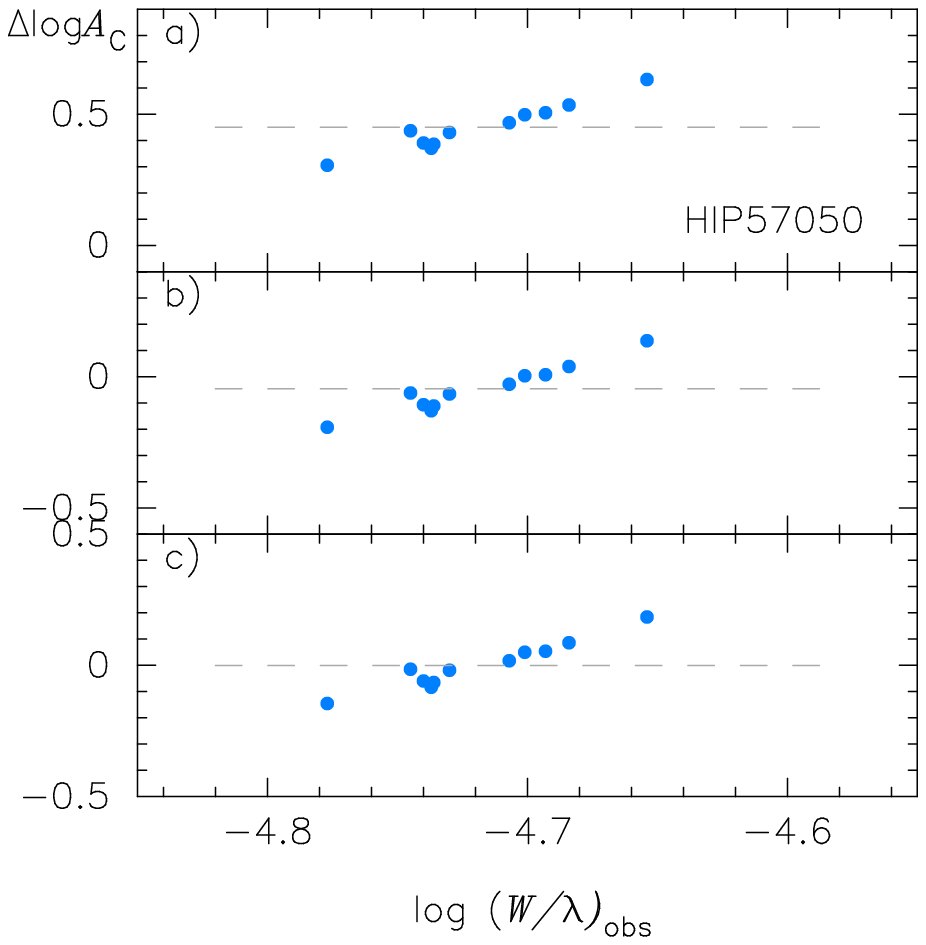}
   \end{center}
   \caption{The resulting logarithmic abundance corrections by the BB analysis 
for the CO lines in HIP\,57050 plotted against the observed values of
log\,${W/\lambda}$. The dashed line shows the mean correction. a) First 
iteration.  b) Second iteration. c) Confirmation of the convergence.
    }
\label{figure6}
\end{figure}

\begin{figure}
   \begin{center}
       \FigureFile(80mm,75mm){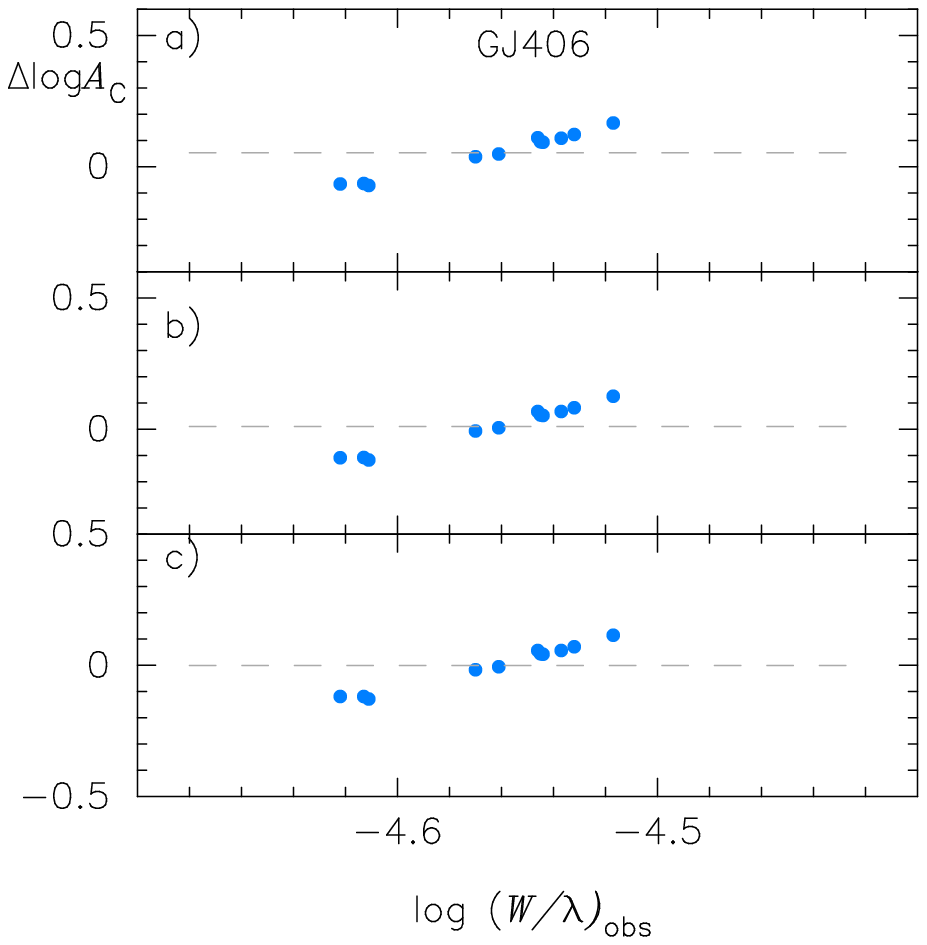}
   \end{center}
   \caption{ The resulting logarithmic abundance corrections by the BB 
analysis for the CO lines in GJ\,406 plotted against the observed values of
log\,${W/\lambda}$. The dashed line shows the mean correction. 
a) First iteration. b) Second iteration. c) Confirmation of the convergence.
    }\label{figure7}
\end{figure}

\begin{figure}
   \begin{center}
       \FigureFile(145mm,95mm){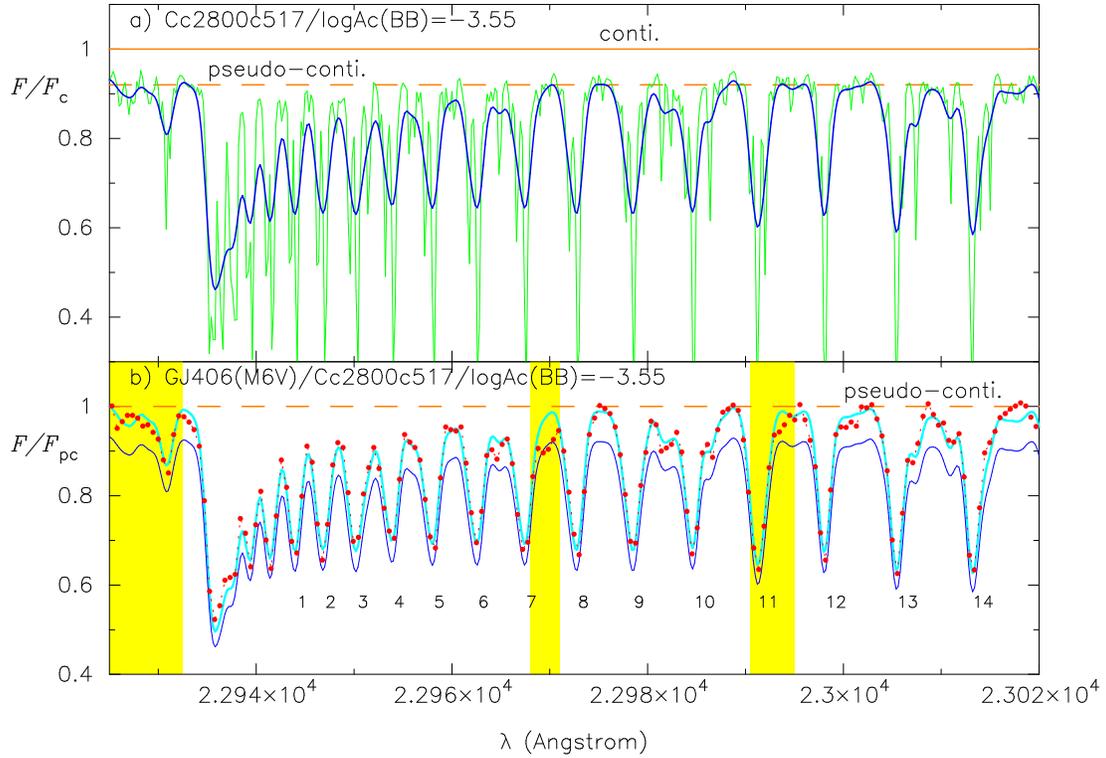}  
   \end{center}
   \caption{a) Theoretical spectra for the model Cc2800c517 and 
with the carbon abundance by the BB analysis,  log\,$A_{\rm C}^{(2)}$ 
(table\,8) in high (thin line) and low (thick line) resolutions. The continuum
level is depressed by about 8\,\% but the pseudo-continuum level is well
defined. b) The theoretical spectrum renormalized by the pseudo-continuum
(thick line) can be matched with the observed spectrum of GJ\,406 (filled 
circles), but the theoretical spectrum normalized by the true continuum 
(copied from figure\,8A above but changed to thin line) cannot. The reference 
numbers of table\,7 are given to the respective CO blends.
 }
\label{figure8}
\end{figure}

\begin{figure}
   \begin{center}
       \FigureFile(145mm,90mm){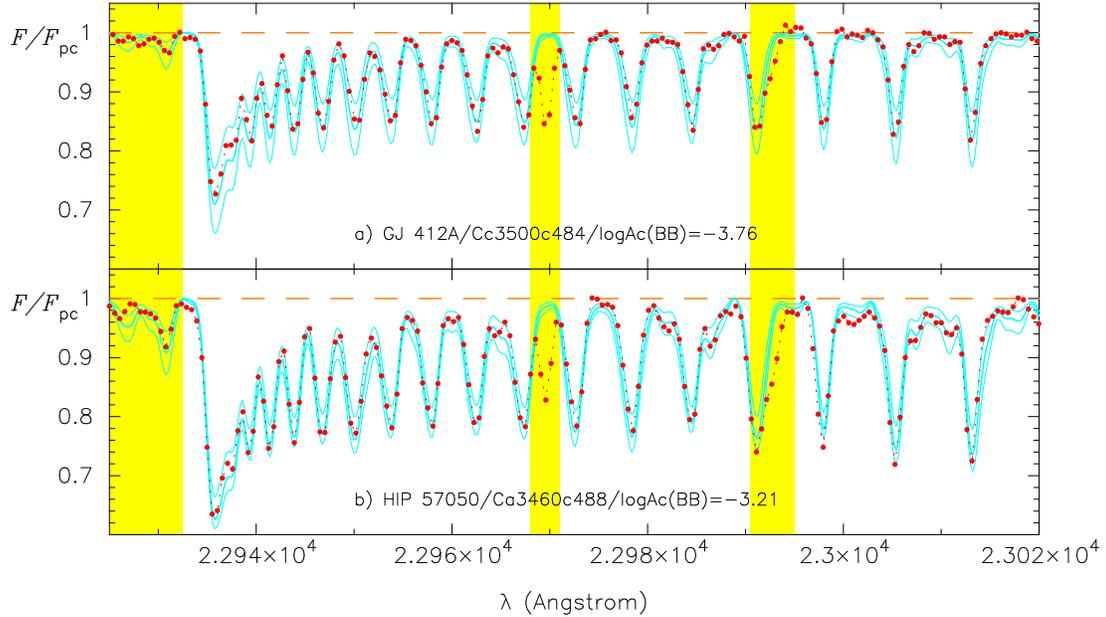}   
   \end{center}
   \caption{
   a) The observed spectrum of GJ\,412A (filled circles) is compared with the 
theoretical one for the model Cc3500c484 (solid lines). The thick line is 
for the carbon abundance log\,$A_{\rm C}^{(2)}$ (table\,8) and the thin 
lines for log\,$A_{\rm C}^{(2)} \pm 0.3$.  
   b) The same as a) but for the observed spectrum of HIP\,57050 (filled
 circles)  compared with the theoretical one for the model Cc3460c488 (solid 
lines). Note that all the spectra are normalized by their pseudo-continua. 
 }
\label{figure9}
\end{figure}

\begin{figure}
   \begin{center}
       \FigureFile(80mm,50mm){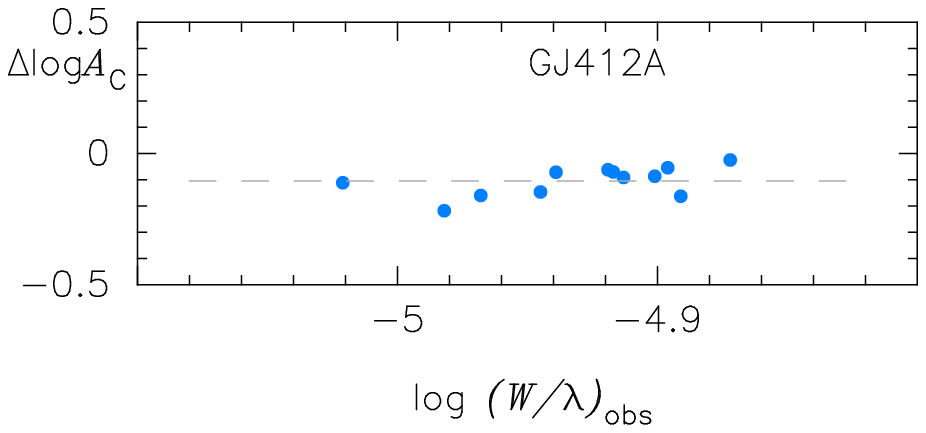}   
   \end{center}
   \caption{ The resulting logarithmic abundance corrections by the BS 
analysis for the CO lines in GJ\,412A plotted against the observed values of
log\,${W/\lambda}$. The dashed line shows the mean correction. }
\label{figure10}
\end{figure}

\begin{figure}
   \begin{center}
       \FigureFile(80mm,50mm){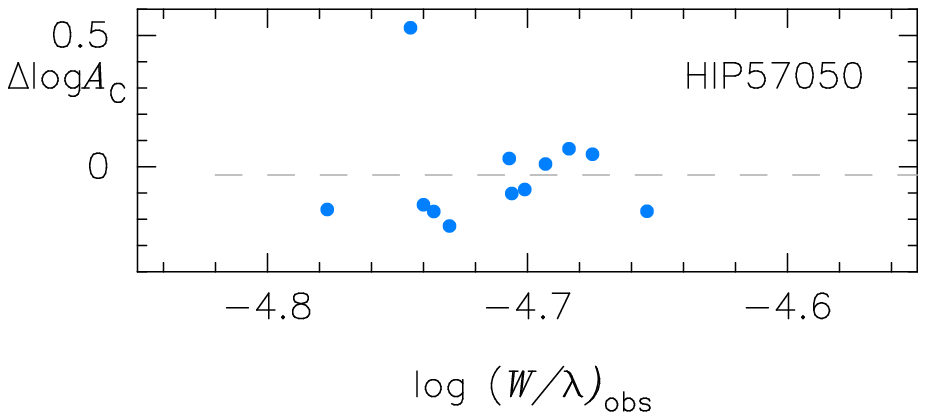}
   \end{center}
   \caption{The resulting logarithmic abundance corrections by the BS 
analysis for the CO lines in HIP\,57050 plotted against the observed values of
log\,${W/\lambda}$. The dashed line shows the mean correction. }
\label{figure11}
\end{figure}

\begin{figure}
   \begin{center}
       \FigureFile(80mm,50mm){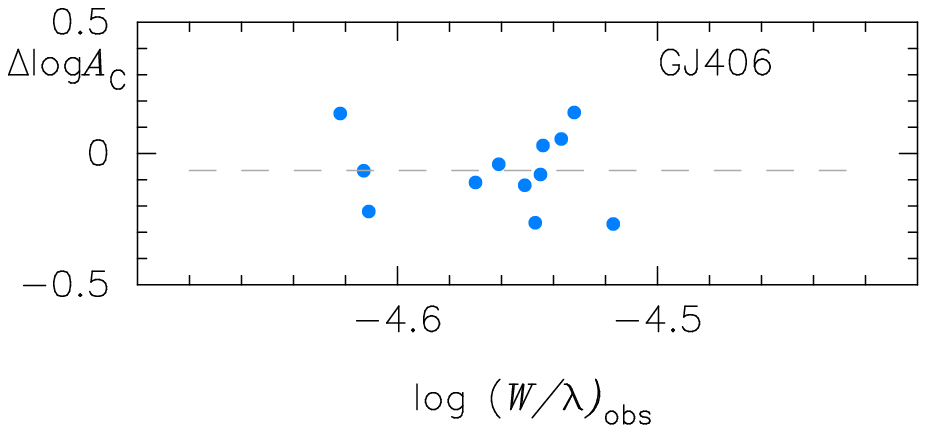}
   \end{center}
  \caption{The resulting logarithmic abundance corrections by the BS 
analysis for the CO lines in GJ\,406 plotted against the observed values of
log\,${W/\lambda}$. The dashed line shows the mean correction. }
\label{figure12}
\end{figure}

\begin{figure}
   \begin{center}
       \FigureFile(160mm,200mm){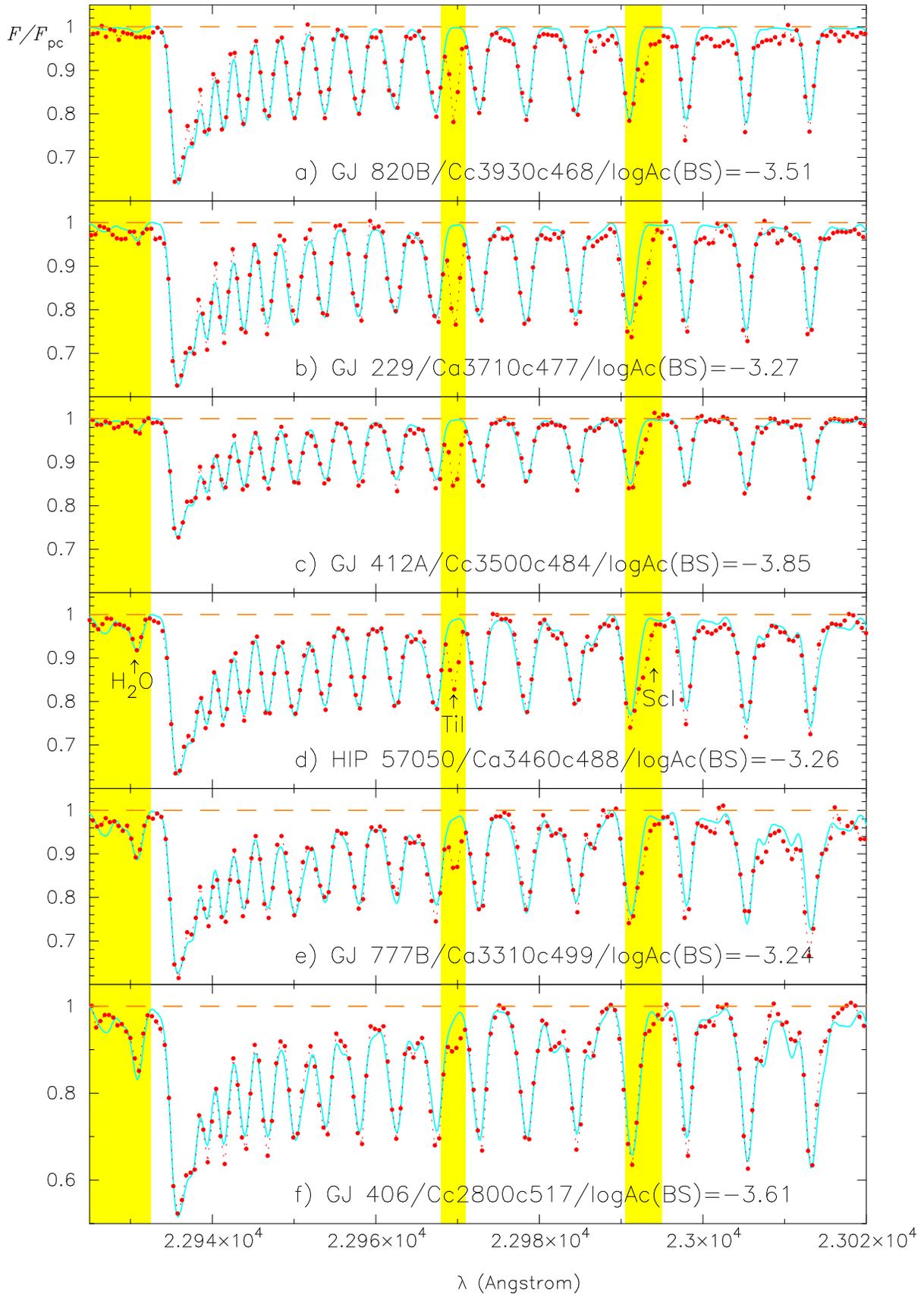}
   \end{center}
   \caption{
    Comparison of observed (filled circles) and predicted (solid lines) 
spectra for the carbon  abundances by the BS method (table\,10) are
 shown for six M dwarfs:
a) GJ\,820B (dM0) . 
b) GJ\,229 (dM2.5).
c) GJ\,412A (dM2).  
d) HIP\,57050 (M4).
e) GJ\,777B (M4.5).
f) GJ\,406 (dM6.5e).
 }
\label{figure13}
\end{figure}

\begin{figure}
   \begin{center}
       \FigureFile(105mm,75mm){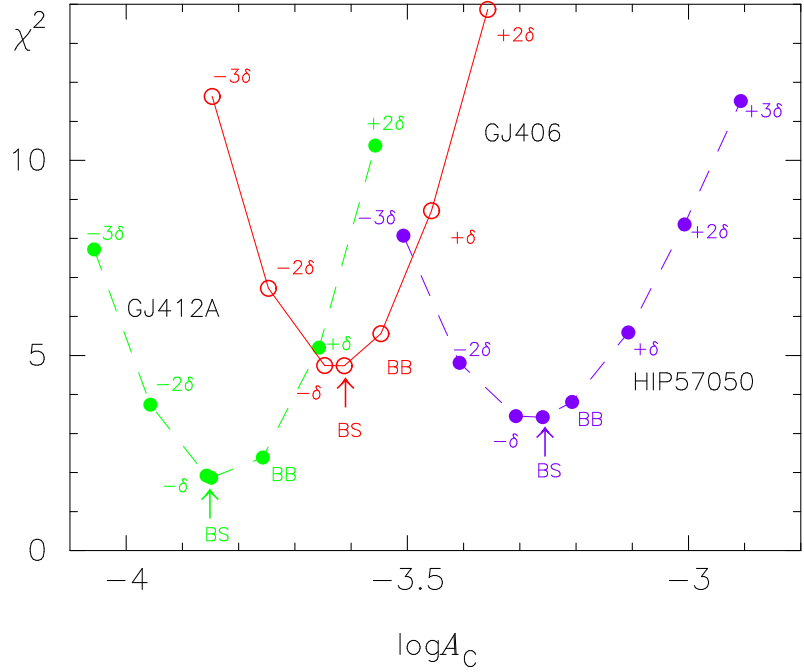}   
   \end{center}
   \caption{$\chi^{2}$ values for the fittings of the observed and
predicted spectra are plotted against log\,$A_{\rm C}$ for GJ\,412A, 
GJ\,406, and HIP\,57050. In this figure, BB and BS correspond to
the carbon abundances based on BB (log\,$A_{\rm C}$(BB)) and BS
(log\,$A_{\rm C}$(BS)) methods, respectively, and
$\pm n\delta$  to log\,$A_{\rm C}{\rm (BB)}\,\pm\,n\delta$ 
($n = 1, 2,$ and 3).  
Note that the results by the BS analysis
are almost at the minima but those by the BB analysis are still not.
    }
\label{figure14}
\end{figure}

\begin{figure}
   \begin{center}
       \FigureFile(175mm,80mm){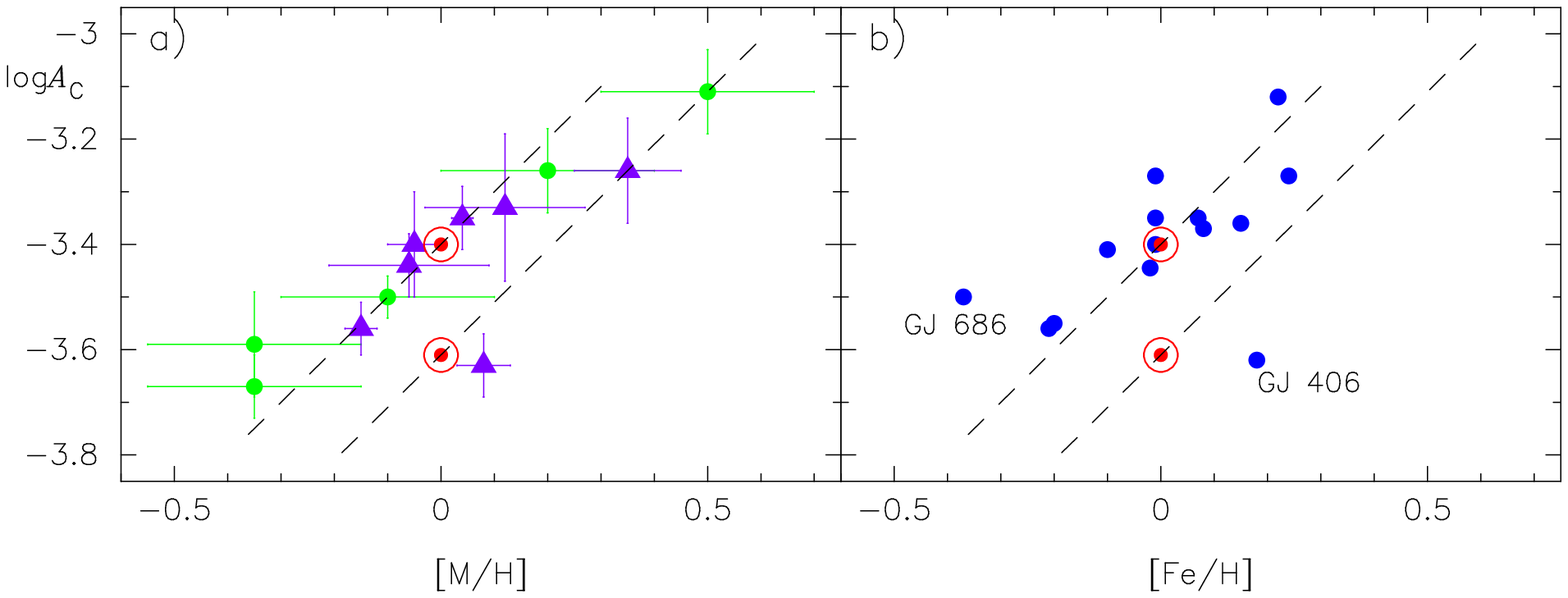}  
   \end{center}
   \caption{ a) Carbon abundances log\,$A_{\rm C}$ (table\,10) are   
compared with the metallicities based on high resolution infrared spectroscopy
by \citet{Mou78} and by \citet{One12}, shown by the filled circles and by
filled triangles, respectively.
b) Carbon abundances log\,$A_{\rm C}$ (table\,10) are compared with the 
[Fe/H] based on the photometric calibrations of [Fe/H] using high 
resolution  spectra by \citet{Nev13}.  The upper dashed line 
represents the locus of C/M (left panel) or of C/Fe (right panel) ratio 
equal to that with the classical high solar carbon abundance indicated
by the upper $\odot$ \citep{Gre91}. 
The lower dashed line represents the locus of C/M (left panel) or of C/Fe 
(right panel) ratio equal to that with the recent downward revised
 solar carbon abundance indicated by the lower $\odot$ \citep{All02}.  }  
\label{figure15}
\end{figure}

\begin{figure}
   \begin{center}
       \FigureFile(80mm,65mm){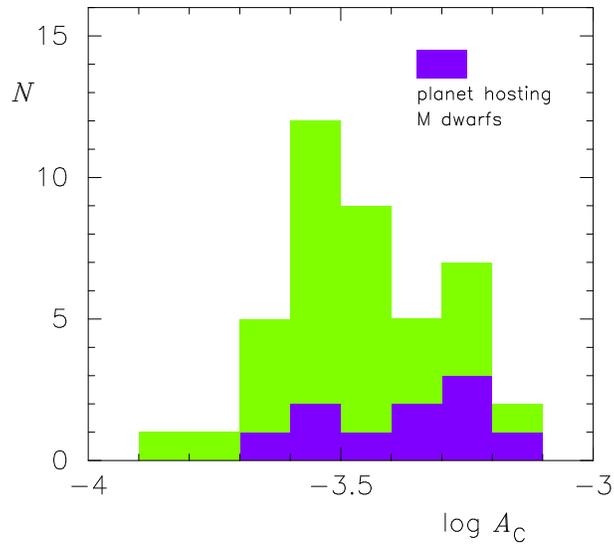}
   \end{center}
   \caption{Frequency distribution of M dwarfs against log\,$A_{\rm C}$.
Note that 10 M dwarfs are  hosting planet(s).    }
\label{figure16}
\end{figure}

\begin{figure}
   \begin{center}
       \FigureFile(160mm,120mm){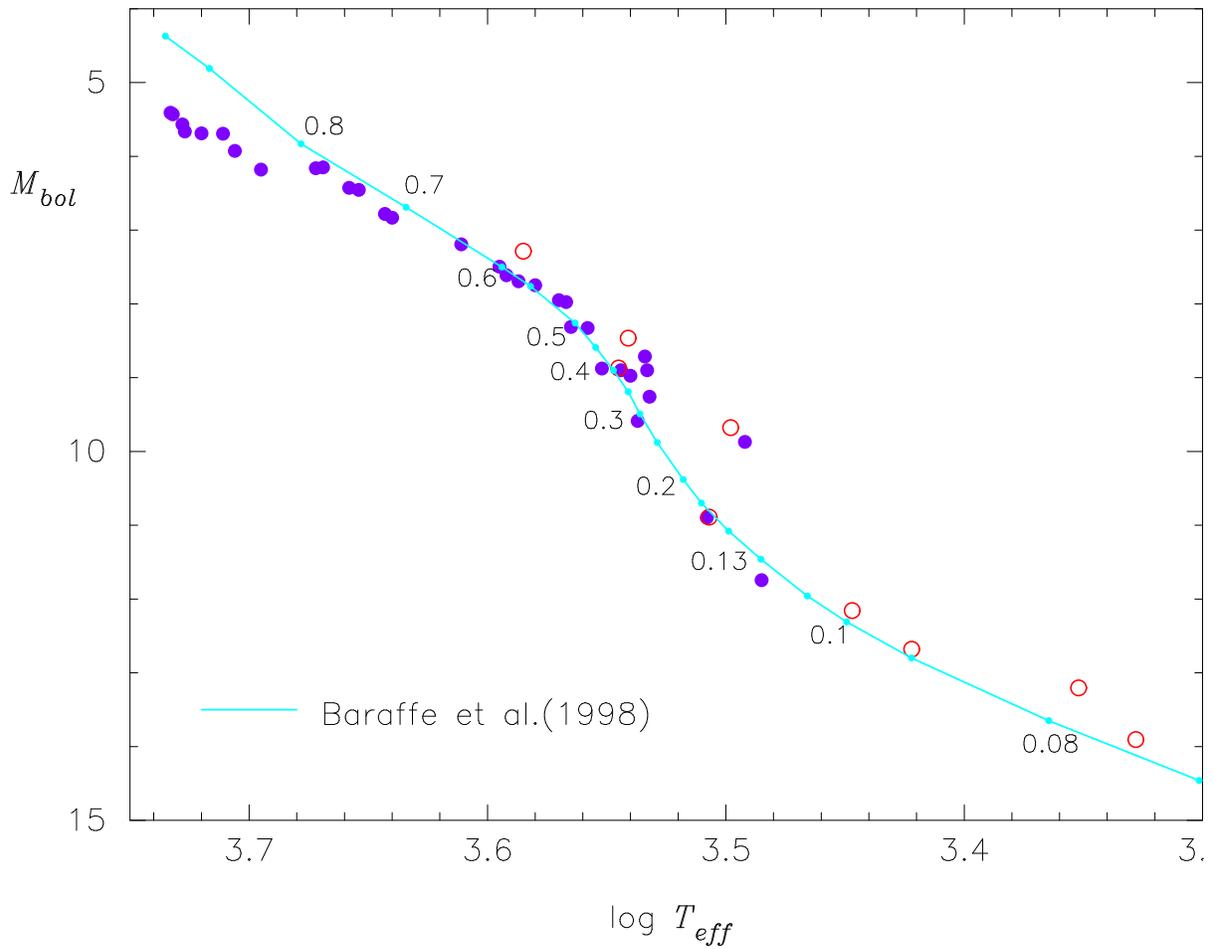}  
   \end{center}
   \caption{The absolute bolometric magnitude  $M_{\rm bol}$ plotted
 against log\,$T_{\rm eff}$ (HR diagram), where $T_{\rm eff}$ values are 
based on the interferometry (filled circles) and infrared flux method 
(open circles) while $M_{\rm bol}$ values from tables\,3 and 4.
The solid line is based on the evolutionary models of the solar metallicity
and the masses of the stellar models are indicated in unit of
$M_{\odot}$ on the curve (as for detail, see \cite{Bar98}).}\label{figure17}
\end{figure}
\clearpage

\begin{longtable}{lrlcrc}
\caption{Target stars. }
\label{tab:first}
\hline
GJ/HIP  &   LHS  & Other name & $K$ mag & SNR    &  Run\footnotemark[$*$]  \\ 
  \hline
\endfirsthead
   \hline
GJ/HIP  &   LHS  & Other name & $K$ mag & SNR    &  Run\footnotemark[$*$]  \\ 
  \hline
\endhead
   \hline
\endfoot
   \hline
\multicolumn{6}{l}{\hbox to 0pt {\parbox{90mm}{\footnotesize
      \footnotemark[$*$] The observing run was either on 2013 May 9 or 
            November 16.

     \footnotemark[$\dagger$] Angular diameter measured by interferometry.

     \footnotemark[$\ddagger$] Planet hosting M dwarf.

}}}
\endlastfoot
   \hline
           &         &            &         &     &       \\
GJ  15A\footnotemark[$\dagger$]   & 3  &  GX And   &    4.02 & 92  &   Nov  \\
GJ 105B    &      16 &   BX Cet  &     6.57 & 85  &   Nov  \\
GJ 166C    &      25 &   DY Eri   &     5.96 &  93 &   Nov  \\
GJ 176\footnotemark[$\ddagger$]  &  196 &   HD 285968 &  5.61 &  104 &  Nov  \\
GJ 179\footnotemark[$\ddagger$]      &   &  G 83$-$37  &  6.94 &  85 & Nov \\
GJ 205\footnotemark[$\dagger$]   &  30   & HD 36395 & 4.04  &  89 &    Nov  \\
GJ 212      &   1775 &  HD 233153 &   5.76 &  89 &   Nov  \\
GJ 229      &   1827 &   HD 42581 &    4.17 &  85 &   Nov    \\
GJ 231.1B  &          &  HD 43587 B &   8.27 &  85 &   Nov \\         
GJ 250B     &  1876 &   HD 50281 B  & 5.72 &  76 &   Nov \\
           &         &            &         &     &       \\
GJ 273       &   33  &  BD+05 1668   & 4.86 &  109 &   Nov     \\
GJ 324B     & 2063  &  $\rho$ Cnc B & 7.67&  90 &   Nov  \\
GJ 338A\footnotemark[$\dagger$]    & 260 & HD 79210 & 3.99 &  112 &  May \\
GJ 338B\footnotemark[$\dagger$]   & 261 &  HD 79211  &  4.14 &  114 &  May \\
GJ 380\footnotemark[$\dagger$] & 280 &  HD 88230  & 3.26 &   66 &   Nov \\
GJ 406       &  36   &  CN Leo        &  6.08 &  84 &    Nov    \\
GJ 411\footnotemark[$\dagger$] & 37  &  HD 95735  &  3.34 &  114 &   May \\
GJ 412A\footnotemark[$\dagger$]  &  38 &   BD+44 2051 & 4.77 &  124 &  May  \\
GJ 436\footnotemark[$\dagger$] \footnotemark[$\ddagger$]  &  310 &  G 121$-$7 
 & 6.07  &  106 & May  \\
GJ 526\footnotemark[$\dagger$] & 47 & HD 119850  &  4.42 &   127 &  May  \\
           &         &            &         &     &       \\
GJ 581\footnotemark[$\dagger$] \footnotemark[$\ddagger$]  & 394  &  HO Lib 
           &  5.84  & 113 &  May  \\
GJ 611B     & 3150 &  HD 144579 B &   9.16 &  106 &   May \\
GJ 649\footnotemark[$\ddagger$] & 3257 &  BD+25 3173  &  5.62  &  97 & May  \\
GJ 686       &  452 &   BD+18 3421 &  5.57  &  68 &     Nov \\
GJ 687\footnotemark[$\dagger$] &  450 & BD+68 732  &  4.55 &  114  &  May  \\
GJ 725A\footnotemark[$\dagger$] &  58 &  HD 173739 &  4.43  &  95   &  May  \\
GJ 725B\footnotemark[$\dagger$] &  59 &  HD 173740 &  5.00  &  113   & May  \\
GJ 768.1C   &       &    GJ 9671C  &  8.01 &   48  &      May  \\
GJ 777B     & 3509 &  HD 190360  &  8.71 &  41   &      May  \\
GJ 783.2B   & 3530 &  HD 191785 B & 8.88 &   111  &    May   \\
           &         &            &         &     &       \\
GJ 797B NE     &       &   G 144$-$26   &   8.17 &  52  &  May  \\ 
GJ 797B SW    &        &              &  8.17  &  52    &   May  \\
GJ 809\footnotemark[$\dagger$] & 3595 &  BD+61 2068 & 4.62 &  93 &  May  \\
GJ 820B\footnotemark[$\dagger$] &   63  &  61 Cyg B  &  2.32 &  86  & Nov \\
GJ 849\footnotemark[$\ddagger$] & 517 &  BD$-$05 5715 &  5.59 & 111  &  Nov  \\
GJ 876\footnotemark[$\ddagger$] &   530  &  IL Aqr &  5.01 &  75  &  Nov \\
GJ 880\footnotemark[$\dagger$]  &  533   &  HD 216899 & 4.52 & 94 &  Nov  \\
GJ 884       &  3885 &   HD 217357 & 4.48 &  79   &   Nov  \\
GJ 3348B   &         &   HD 35956 B   &  8.79 &  38  &   Nov \\
HIP 12961\footnotemark[$\ddagger$]  & & CD$-$23 1056 &  6.74 &  70  &  Nov  \\
           &         &            &         &     &       \\
HIP 57050\footnotemark[$\ddagger$] & 2443 &  GJ 1148  &  6.82 & 112  &  May  \\
HIP 79431\footnotemark[$\ddagger$] & &  LP 804$-$27  &  6.59   & 112  & May \\
           &         &            &         &     &       \\
\hline 
\end{longtable}

\begin{table}
\begin{center}
\caption{Echelle setting}
\label{table-echelle}
\begin{tabular}{ccc}
\hline
  Order  &  Wavelength range (\AA) & Dispersion ({\AA} pixel$^{-1}$) \\ 
\hline           
29th  & 19397$-$19867 & 0.460  \\
28th  & 20089$-$20575 & 0.475  \\
27th  & 20833$-$21335  &  0.491  \\  
26th  & 21634$-$22154 & 0.508  \\ 
25th  & 22500$-$23039 & 0.527  \\ 
24th  & 23437$-$23997 & 0.547  \\  
23rd  & 24456$-$25038 & 0.569  \\
\hline 
\end{tabular}
\end{center}
\end{table}

\begin{longtable}{llcccccccc}
  \caption{Fundamental parameters based on $T_{\rm eff}$ by
 the interferometry}
\label{tab:second}
   \hline
obj. & sp.ty.\footnotemark[$*$] & $p$(msec)\footnotemark[$\dagger$] & 
$F_{3.4}$(mag)\footnotemark[$\ddagger$]  & $M_{3.4}$(mag)\footnotemark[$\S$]
 & $T_{\rm eff}$ & $M_{\rm bol}$(mag)\footnotemark[$\|$] & $R/R_{\odot}$ & 
$M/M_{\odot}$ & log\,$g$   \\ 
   \hline
\endfirsthead
   \hline
obj. & sp.ty.\footnotemark[$*$] & $p$(msec)\footnotemark[$\dagger$] & 
$F_{3.4}$(mag)\footnotemark[$\ddagger$]  & $M_{3.4}$(mag)\footnotemark[$\S$]
 & $T_{\rm eff}$ & $M_{\rm bol}$(mag)\footnotemark[$\|$] & $R/R_{0}$ & 
$M/M_{0}$ & log\,$g$  \\ 
   \hline
\endhead
   \hline
\endfoot
   \hline
\multicolumn{9}{l}{\hbox to 0pt {\parbox{150mm}{\footnotesize
  \footnotemark[$*$] The spectral types  for
M dwarfs we analyze in this paper are by \citet{Joy74} and  those for
other objects from \citet{Boy12}.

\footnotemark[$\dagger$] Parallax by Hipparcos \citep{Lee07}.

\footnotemark[$\ddagger$] $WISE$ $W1$ band centered at 3.4\,$\mu$m
 \citep{Wri10}.

\footnotemark[$\S$] Absolute magnitude at 3.4\,$\mu$m. 

\footnotemark[$\|$] Absolute bolometric magnitude based on 
 $L/L_{\odot}$ of Table 6 in \citet{Boy12}.

\footnotemark[$\#$] We apply different $T_{\rm eff}$ values - low and high -
for this object,  and L implies that this is the case of low $T_{\rm eff}$
(see subsection\,3.1). 
}}}
\endlastfoot
   \hline
       &       &                   &                   &                 &      &        &        &     &  \\
GJ~15A & dM2.5 & 278.76 $\pm$~0.77 & 3.853 $\pm$~0.099 & 6.08 $\pm$~0.10 & 
3563 $\pm$~11 & 8.88   & 0.386 &  0.423 & 4.890 \\
GJ~ 33 & K2V & 134.14 $\pm$~0.51 & 3.561 $\pm$~0.142 & 4.20 $\pm$~0.15 &
 4950 $\pm$~14   & 6.18   &  &  & \\ 
GJ~53A & G5Vp & 132.38 $\pm$~0.82 & 3.326 $\pm$~0.141 & 3.94 $\pm$~0.15 &
 5348 $\pm$~26  & 5.57   &  &  & \\ 
GJ~ 75 & K0V &  99.33 $\pm$~0.53 &  3.753$\pm$~0.091 & 3.74 $\pm$~0.10 & 
5398 $\pm$~75  & 5.43  &  &  &  \\
GJ~105 & K3V & 139.27 $\pm$~0.45 & 3.395 $\pm$~0.127 & 4.11 $\pm$~0.13 & 
4662 $\pm$~17 & 6.15    &  &  &  \\
GJ~144 & K2V & 310.94 $\pm$~0.16 & 2.970 $\pm$~0.251 & 5.43 $\pm$~0.25 & 
 5077 $\pm$~35 & 5.93  
 &  &  &  \\
GJ~166A & K1Ve & 200.62$\pm$~0.23 & 2.463$\pm$~0.242 & 3.97 $\pm$~0.24 & 
 5143 $\pm$~14 & 5.69  &  &  &  \\
GJ~205 & dM3 & 176.77 $\pm$~1.18 & 3.743 $\pm$~0.120 & 4.98 $\pm$~0.13 & 
 3801 $\pm$~ 9 &  7.75  & 0.574 & 0.615  & 4.710 \\
GJ~338A & dM0.5 & 161.59 $\pm$~5.23   &                   &                 & 
3907 $\pm$~35 & 7.61 & 0.577  & 0.622  & 4.709  \\
GJ~338B & dM0.5 & 159.48 $\pm$~6.61   &                   &                 & 
 3867 $\pm$~37 & 7.69  &  0.567   & 0.600  & 4.709  \\ 
       &       &                   &                   &                 &      &        &        &     &  \\
GJ~380 & dM0.5 & 205.21 $\pm$~0.54 & 3.076 $\pm$~0.149 & 4.64 $\pm$~0.15 & 
 4081 $\pm$~15 & 7.19  &   0.642 & 0.660  & 4.643  \\
GJ~411 & dM2 & 392.64 $\pm$~0.67 & 3.239 $\pm$~0.136 & 6.21 $\pm$~0.14 & 
  3465 $\pm$~17 & 8.97   &  0.392 & 0.403  & 4.857  \\
GJ~412A & dM2 & 206.27 $\pm$~1.00 & 4.638 $\pm$~0.085 & 6.21 $\pm$~0.10 &
 3497 $\pm$~39 & 8.90 &  0.398 & 0.403  & 4.843  \\
GJ~436 & dM3.5 & 98.61 $\pm$~2.33 & 5.987 $\pm$~0.052 & 5.96 $\pm$~0.10 & 
 3416 $\pm$~53 & 8.71  &  0.455 & 0.472  & 4.797  \\
GJ~526 & dM3 & 185.49 $\pm$~1.10 & 4.372 $\pm$~0.095 & 5.71 $\pm$~0.11 & 
 3618 $\pm$~31 & 8.33  & 0.484  & 0.520 & 4.784  \\
GJ~551 & M5.5V & 771.64 $\pm$~2.60 & 4.195 $\pm$~0.086 & 8.63 $\pm$~0.09 & 
 3054 $\pm$~79 & 11.74  &  &  &  \\
GJ~570A & K4V & 171.22 $\pm$~0.94 & 3.159 $\pm$~0.012 & 4.33 $\pm$~0.02 &
  4507 $\pm$~58 & 6.46  &  &  &  \\
GJ~581 & dM4 & 160.91 $\pm$~2.62 & 5.694 $\pm$~0.055 & 6.73 $\pm$~0.09 & 
 3442 $\pm$~54 & 9.59  & 0.299  & 0.297  & 4.959 \\
GJ~631 & K0V & 102.55 $\pm$~0.40 & 3.797 $\pm$~0.110 & 3.85 $\pm$~0.12 & 
 5337 $\pm$~41 & 5.66  &   &   &  \\
GJ~687 & dM4 & 220.84 $\pm$~0.94 & 4.397 $\pm$~0.094 & 6.12 $\pm$~0.10 &
 3413 $\pm$~28 & 8.90  & 0.418  & 0.413  & 4.811 \\
       &       &                   &                   &                 &      &        &        &     &  \\
GJ~699 & M4V & 548.31 $\pm$~1.51 & 4.386 $\pm$~0.073 & 8.08 $\pm$~0.08 & 
 3224 $\pm$~10 & 10.90  &  &   &  \\
GJ~702A & K0VE & 196.62 $\pm$~1.38 &  &  &  5407 $\pm$~52 & 5.41 
   &  &  &  \\
GJ~702B & K5Ve & 196.62 $\pm$~1.38 &  &  &  4393 $\pm$~149 & \ 6.78 
    &  &  &  \\
GJ~725A & dM4 & 280.18 $\pm$~2.18 & 4.498 $\pm$~0.226 & 6.74 $\pm$~0.24 & 
 3407 $\pm$~15 & 9.26   & 0.356 & 0.318  & 4.837 \\
GJ~725B-L\footnotemark[$\#$] & dM4.5 & 289.48 $\pm$~3.21 & 5.014 $\pm$~0.325 & 7.32 $\pm$~0.35 & 3104 $\pm$~28 & 9.87   & 0.323 & 0.235 & 4.790 \\
GJ~764 & K0V & 173.77 $\pm$~0.18 & 2.786 $\pm$~0.129 & 3.99 $\pm$~0.13 & 
 5246 $\pm$~26 & 5.69  &  &  &  \\
GJ~809 & dM2 & 141.87 $\pm$~0.64 & 4.501 $\pm$~0.088 & 5.26 $\pm$~0.10 &
 3692 $\pm$~22 &7.98   & 0.547 & 0.573 & 4.720 \\
GJ~820A & K5V & 287.13 $\pm$~1.51  &                   &                & 
 4361 $\pm$~17 & 6.83    &    &    &   \\
GJ~820B & dM0 & 285.42 $\pm$~0.72  &                   &                & 
  3932 $\pm$~25 & 7.50   & 0.601  & 0.629 & 4.679 \\
GJ~845 & K5V & 276.06 $\pm$~0.28 & 2.899 $\pm$~0.192  & 5.10 $\pm$~0.19
 &  4555 $\pm$~24 &  &  
  &  &  \\
       &       &                   &                   &                 &      &        &        &    &  \\
GJ~880 & dM2.5 & 146.09 $\pm$~1.00 & 4.432 $\pm$~0.080 & 5.26 $\pm$~0.09 & 
3713 $\pm$~11 & 7.95  & 0.642 & 0.660  & 4.643 \\
GJ~887 & M0.5V & 305.26 $\pm$~0.70 & 3.243 $\pm$~0.121 & 5.67 $\pm$~0.13&
 3676 $\pm$~35 & 8.31  &  &  &  \\
GJ~892 & K3V & 152.76 $\pm$~0.29 & 3.346 $\pm$~0.098 & 4.27 $\pm$~0.10 & 
4699 $\pm$~16 & 6.16   &  &  &  \\
       &       &                   &                   &                 &      &        &        &   &    \\
\hline
\end{longtable}

\begin{longtable}{llcccccccc}
  \caption{Fundamental parameters based on $T_{\rm eff}$ by the infrared
 flux method}
\label{tab:third}
   \hline
obj. & sp.ty.\footnotemark[$*$] & $p$(msec)\footnotemark[$\dagger$] & 
$F_{3.4}$(mag)\footnotemark[$\ddagger$]  & $M_{3.4}$(mag)\footnotemark[$\S$]
 & $T_{\rm eff}$ & $M_{\rm bol}$(mag)\footnotemark[$\|$] &
$R/R_{\odot}$ & $M/M_{\odot}$ & log\,$g$  \\ 
   \hline
\endfirsthead
   \hline
obj. & sp.ty.\footnotemark[$*$] & $p$(msec)\footnotemark[$\dagger$] & 
$F_{3.4}$(mag)\footnotemark[$\ddagger$]  & $M_{3.4}$(mag)\footnotemark[$\S$] 
& $T_{\rm eff}$ & $M_{\rm bol}$(mag)\footnotemark[$\|$] &
$R/R_{\odot}$ & $M/M_{\odot}$ & log\,$g$  \\ 
   \hline
\endhead
   \hline
\endfoot
   \hline
\multicolumn{9}{l}{\hbox to 0pt {\parbox{150mm}{\footnotesize

  \footnotemark[$*$ ] The spectral types  for
M dwarfs we analyze in this paper are by \citet{Joy74} and  those for
other objects from  \citet{Tsu96}.

\footnotemark[$\dagger$ ] Parallax by Hipparcos \citep{Lee07}, except for
GJ\,406 and GJ\,644C by RECONS and GJ\,3849 by \citet{Gli91}.

\footnotemark[$\ddagger$ ] $WISE$ $W1$ band centered at 3.4\,$\mu$m
 \citep{Wri10}.

\footnotemark[$\S$ ] Absolute magnitude at 3.4\,$\mu$m.

\footnotemark[$\|$ ] Absolute bolometric magnitude based on     
log\,$f_{\rm bol}$(erg\,cm$^{-2}$\,sec$^{-1}$)  derived 
by integration of the observed SED  in Table\,1 of \citet{Tsu96} and 
the parallax.

\footnotemark[$\#$ ] We apply different $T_{\rm eff}$ values - low and high -
for this object,  and L implies that this is the case of low $T_{\rm eff}$
(see subsection\,3.1).

 \footnotemark[$**$ ]
The empirical formulae  by \citet{Boy12} are inapplicable to such a cool
M dwarf, and mass is estimated by the use of the mass-luminosity ($M_{\rm K}$)
relation by \citet{Del00} with $K = 6.08$ \citep{Gez87}. Radius is
inferred from $T_{\rm eff}$ and $L_{\rm bol} = 1.06\times10^{-3}\,L_{\odot} $ 
based on $M_{\rm bol}$.

  \footnotemark[$***$]  Compared with $T_{\rm eff}$ = 3465\,K by the
interferometry (table 3), $T_{\rm eff}$ by the infrared flux method is higher
 by 45\,K. 

  \footnotemark[$****$]  Compared with $T_{\rm eff}$ = 3224\,K by the
interferometry (table 3), $T_{\rm eff}$ by the infrared flux method is lower 
by 14\,K. 
}}}
\endlastfoot
   \hline
       &       &                   &                   &                 &      &        &        &    &   \\
GJ~273-L\footnotemark[$\#$] & dM4 & 262.98 $\pm$~1.39 & 4.723 $\pm$~0.074 &
 6.82 $\pm$~0.09 & 3150 & ~9.68   & 0.141 & 0.079  & 5.042 \\
GJ~406\footnotemark[$**$] & dM6.5e & 419.10 $\pm$~2.10 & 5.807 $\pm$~0.055 & 
8.92 $\pm$~0.07 & 2800 &  12.16 & 0.139  &  0.103  & 5.166  \\
GJ~411\footnotemark[$***$] & M2V & 392.64 $\pm$~0.67 & 3.239 $\pm$~0.136 & 
6.21 $\pm$~0.14 & 3510 & ~8.87  &  & & \\
GJ~644C & M7V & 148.92 $\pm$~4.00 & 8.588 $\pm$~0.023 & 9.45 $\pm$~0.08 & 
2640 & 12.68 &  & & \\
GJ~699\footnotemark[$****$] & M4V & 548.31 $\pm$~1.51 & 4.386 $\pm$~0.073 & 
8.08 $\pm$~0.08 & 3210 &  10.89  & & &  \\
GJ~752A & M3V & 170.36 $\pm$~1.00 & 4.466 $\pm$~0.078 & 5.62 $\pm$~0.09 & 
 3475 & ~8.47  & & & \\
GJ~752B & M8V & 170.36 $\pm$~1.00 & 8.465 $\pm$~0.023 & 9.62 $\pm$~0.04 & 
 2250  & 13.21  & & & \\
GJ~884 & dM0.5 & 121.69 $\pm$~0.69 & 4.424 $\pm$~0.087 & 4.85 $\pm$~0.10 & 
 3850 & ~7.29  &  0.551  &  0.581 & 4.720 \\
GJ~3849 & M9V & 95.00 $\pm$~5.70 & 10.431 $\pm$~0.023 & 10.32 $\pm$~0.15 &
 2130 & 13.91  & & &  \\
       &       &                   &                   &                 &  
  &     &        &        &     \\
\hline
\end{longtable}

\begin{longtable}{llccccccc}
  \caption{Fundamental parameters based on $T_{\rm eff}$ by the $M_{3.4} -
 {\rm log}\,T_{\rm eff}$ relation (see figure1)}
\label{tab:fourth}
\hline
obj. & sp.ty.\footnotemark[$*$] & $p$(msec)\footnotemark[$\dagger$] &
$F_{3.4}$(mag)\footnotemark[$\ddagger$]  & $M_{3.4}$(mag)\footnotemark[$\S$]
 & $T_{\rm eff}$ & $R/R_{\odot}$ & $M/M_{\odot}$ & log\,$g$  \\
   \hline
\endfirsthead
   \hline
obj. & sp.ty.\footnotemark[$*$] & $p$(msec)\footnotemark[$\dagger$] &
$F_{3.4}$(mag)\footnotemark[$\ddagger$]  & $M_{3.4}$(mag)\footnotemark[$\S$]
 & $T_{\rm eff}$ & $R/R_{\odot}$ & $M/M_{\odot}$ & log\,$g$  \\
   \hline
\endhead
   \hline
\endfoot
   \hline
\multicolumn{9}{l}{\hbox to 0pt {\parbox{150mm}{\footnotesize
  \footnotemark[$*$] The spectral types beginning with dM are by \citet{Joy74}
and  those beginning with M by SIMBAD.

\footnotemark[$\dagger$] Parallax by Hipparcos \citep{Lee07}.

\footnotemark[$\ddagger$] $WISE$ $W1$ band centered at 3.4\,$\mu$m
 \citep{Wri10}.

\footnotemark[$\S$] Absolute magnitude at 3.4\,$\mu$m.

\footnotemark[$\|$] We apply different $T_{\rm eff}$ values - low and high -
for this object,  and H implies that this is the case of high $T_{\rm eff}$
(see subsection\,3.1).

\footnotemark[$\#$] $F_{3.4}$ observed for GJ\,797B is 7.213\,mag, but this is a sum of two components of
about equal brightness. The values shown in 4-th and 5-th columns are for a component.

}}}
\endlastfoot
   \hline
       &       &                   &                   &                 &      &        &        &     \\
GJ~105B  &  dM4.5 &  139.270 $\pm$~0.450 &  6.449 $\pm$~0.044  &  7.17 $\pm$~0.05 &  3360 &   0.305 &  0.304 &  4.954 \\
GJ~166C &  dM4e &   200.620 $\pm$~0.230 &  5.806 $\pm$~0.045  &  7.32 $\pm$~0.05 &  3337 &  0.289 &  0.284 &  4.972 \\
GJ~176   &  dM2.5e  & 107.830 $\pm$~2.850  & 5.434 $\pm$~0.066  &  5.60 $\pm$~0.12 &  3616 &  0.454 &  0.479 &  4.804 \\
GJ~179 & dM3.5e   &   81.380 $\pm$~4.040  & 6.785 $\pm$~0.034  &  6.34 $\pm$~0.14 &  3476 &  0.378 &  0.393 &  4.877 \\
GJ~212  & dM1  &  80.400 $\pm$~1.690 &  5.586 $\pm$~0.049  &  5.11 $\pm$~0.09 &  3757 &  0.517 & 0.546 & 4.748 \\
GJ~229  & dM2.5 & 173.810 $\pm$~0.990 &  4.060 $\pm$~0.128 &   5.26 $\pm$~0.14 &  3707 &   0.496 &  0.524 &  4.766 \\
GJ~231.1B & M3.5  &  51.950 $\pm$~0.400 &  8.044 $\pm$~0.023  &  6.62 $\pm$~0.04  & 3442 &  0.358 &  0.369 &  4.897 \\
GJ~250B & M2  & 114.810 $\pm$~0.440 &  5.497 $\pm$~0.052  &  5.80 $\pm$~0.06  & 3567 &  0.430 &  0.452 &  4.827 \\
GJ~273-H\footnotemark[$\|$]   & dM4   & 262.980 $\pm$~1.390  & 4.723 $\pm$~0.074  & 6.82 $\pm$~0.09 &   3415 &  0.341 &  0.349 &  4.915 \\
GJ~324B  & M4 &   81.030 $\pm$~0.750 &  7.483 $\pm$~0.025  &  7.03 $\pm$~0.05 &  3382 &  0.319 &  0.323 &  4.938 \\
       &       &                   &                   &                 &      &        &        &     \\
GJ~611B & M4   &   68.870 $\pm$~0.330 &  8.971 $\pm$~0.023  &  8.16 $\pm$~0.03 &  3202 &  0.185 &  0.145 &  5.063 \\
GJ~649 & dM2  &     96.670 $\pm$~1.390 &  5.502 $\pm$~0.065 &   5.43 $\pm$~0.10 &  3660 &  0.475 &  0.501 &  4.784 \\
GJ~686 & dM1  &   123.670 $\pm$~1.610  & 5.479 $\pm$~0.070 &  5.94 $\pm$~0.10 &  3538 &  0.414 &  0.434 &  4.842 \\
GJ~725B-H\footnotemark[$\|$]  &  dM4.5   &  289.480 $\pm$~3.210 &  5.014 $\pm$~0.325  &  7.32 $\pm$~0.35 &  3337 &  0.288 &  0.284  & 4.972 \\
GJ~768.1C & -  &  52.110 $\pm$~0.290  & 7.801 $\pm$~0.026  &  6.39 $\pm$~0.04 &  3470 &  0.375 &  0.389 &  4.880 \\
GJ~777B & M4.5  &    63.060 $\pm$~0.340 &  8.495 $\pm$~0.024 &  7.49 $\pm$~0.04 &  3310 &  0.269 &  0.259 &  4.991 \\
GJ~783.2B & M4   &   49.040 $\pm$~0.650 &  8.670 $\pm$~0.022   & 7.12 $\pm$~0.05 &  3368 &  0.310 &  0.311 &  4.949 \\
GJ~797B-NE\footnotemark[$\#$] & M2.5 &  47.740 $\pm$~0.480  & 7.966 $\pm$~0.027  &  6.36 $\pm$~0.05 &  3473 &  0.377 &  0.391 &  4.878 \\
GJ~797B-SW\footnotemark[$\#$] & M2.5 &  47.740 $\pm$~0.480  & 7.966 $\pm$~0.027  &  6.36 $\pm$~0.05 &  3473 &  0.377 &  0.391 &  4.878 \\
GJ~849  & dM3.5  &  109.940 $\pm$~2.070  & 5.545 $\pm$~0.056  &  5.75 $\pm$~0.10 &  3580 &  0.436 &  0.459 &  4.821 \\
       &       &                   &                   &                 &      &        &        &     \\
GJ~876  &  dM4.5  &  213.280 $\pm$~2.120  & 4.844 $\pm$~0.077  &  6.49 $\pm$~0.10 &  3458 &  0.368 &  0.381 &  4.888 \\
GJ~3348B  & M4   &  35.500 $\pm$~0.950 &  8.585 $\pm$~0.023  & 6.34 $\pm$~0.08 &  3476 &  0.379 &  0.393 &  4.876 \\
HIP~12961 &  M0  &  43.450 $\pm$~1.720   & 6.619 $\pm$~0.039  & 4.81 $\pm$~0.12  & 3890 &  0.565 &  0.595  & 4.709 \\
HIP~57050 & M4   &   90.060 $\pm$~2.750 &  6.665 $\pm$~0.041  &  6.44 $\pm$~0.11 &  3464 &  0.371 &  0.385 &  4.884 \\
HIP~79431 & M3  &    69.460 $\pm$~3.120  &  6.487 $\pm$~0.042 &  5.70 $\pm$~0.14 &  3592 &  0.442 &  0.466 &  4.815 \\
       &       &                   &                   &                 &      &        &        &     \\
\hline
\end{longtable}

\begin{table}
  \begin{center}
  \caption{Effect of the collision partners on the collision half-width for 
CO pure rotational transitions (see equation 1 as for the definitions
of $\gamma_{0}$ and $n$, under which the collision partner is shown in
parenthesis)} 
\label{tab:fifth}
    \begin{tabular}{cccccc}
      \hline
  $ J^{'} - J^{''}$  & $\nu$   & $\gamma_{0}$  &
$\gamma_{0}$  & $n$ & $n$  \\
                     &  (cm$^{-1})$ & (H$_2$)  &
  (air)  &  (H$_2$)  &  (air)  \\
      \hline
       1 - 0 & ~~3.845 &  0.0739 & 0.0797 & 0.574 & 0.76 \\
      10 - 9 & ~38.426 & 0.0721 & 0.0580 & 0.628 & 0.75 \\
      20 -19 & ~76.705 & 0.0700 & 0.0510 & 0.603 & 0.67 \\
      30 -29 & 114.691 & 0.0681 & 0.0436 & 0.594 & 0.67 \\
      \hline
    \end{tabular}
  \end{center}
\end{table}

\begin{table}
  \caption{ Spectroscopic data of CO lines  (2-0 band) } 
\label{tab:6-th}
  \begin{center}
    \begin{tabular}{ccccc}
      \hline
      \smallskip
  Ref. no.\footnotemark[$*$] & wavelength\footnotemark[$\dagger$] & log\,$gf$ 
& L.E.P. & Rot. ID.  \\
     &  (\AA)     &            & (cm$^{-1}$) &      \\
      \hline
    1 &  22943.492 &   -5.130 & 6289.187 &  R 57 \\
      &  22944.160 &   -5.279 & 3782.656 &  R 44 \\
    2 &  22946.312 &   -5.120 & 6507.445 &  R 58 \\
      &  22947.055 &   -5.292 & 3615.553 &  R 43 \\
    3 &  22949.546 &   -5.109 & 6729.297 &  R 59 \\
      &  22950.354  &  -5.305 & 3452.157 &   R 42 \\
    4 &  22953.196 &   -5.099 & 6954.738 &  R 60 \\
      &  22954.055 &   -5.318 & 3292.474 &  R 41 \\
    5 &  22957.262 &   -5.089 & 7183.758 &  R 61 \\
      &  22958.159 &   -5.332 & 3136.510 &  R 40 \\
    6 &  22961.747 &   -5.079 & 7416.348 &  R 62 \\
      &  22962.665  &  -5.345 & 2984.271 &  R 39 \\
    7 &  22966.652 &   -5.069 & 7652.496 &  R 63 \\
      &  22967.572  &  -5.359 & 2835.763 &  R 38 \\
    8 &  22971.977 &   -5.059 & 7892.199 &  R 64 \\
      &  22972.882  &  -5.374 & 2690.992 &  R 37 \\
    9 &  22977.724 &   -5.050 & 8135.441 &  R 65 \\
      &  22978.592  &  -5.388 & 2549.963 &  R 36 \\
   10 &  22983.895 &   -5.040 & 8382.219 &  R 66 \\
      &  22984.704  &  -5.403 & 2412.681 &  R 35 \\
   11\footnotemark[$\ddagger$]  &  22990.491 &   -5.031 & 8632.520 &  R 67 \\
      &  22991.217 &   -5.418 & 2279.152 &  R 34 \\
   12 &  22997.514 &   -5.021 & 8886.336 &  R 68 \\
      &  22998.131 &   -5.434 & 2149.381 &  R 33 \\
   13 &  23004.965 &   -5.012 & 9143.652 &  R 69 \\
      &  23005.446 &   -5.449 & 2023.372 &  R 32 \\
   14 &  23012.845  &  -5.003 & 9404.465 &  R 70 \\
      &  23013.162 &   -5.466 & 1901.131 &  R 31 \\ 
     \hline
  \multicolumn{5}{l}{\hbox to 0pt{\parbox{80mm}{\footnotesize
     \par\noindent
     \footnotemark[$*$] Ref.\,no. refers to figure 4.
     \par\noindent
     \footnotemark[$\dagger$] In vacuum.
     \par\noindent
     \footnotemark[$\ddagger$] Blended with Sc\,I line and not used in 
          our analysis. 
     }}}
   \end{tabular}
  \end{center}
   \end{table}

\begin{longtable}{llllcclccc}
  \caption{Carbon abundances in M dwarfs - a preliminary result by the BB
 method}
\label{tab:7-th}
     \hline
obj. &  $T_{\rm eff}$ & log\,$g$ &  model 1\footnotemark[$*$] & ${\Delta\,{\rm
 log}\,A_{\rm C}^{(1)}}$\footnotemark[$\dagger$] & log\,$A_{\rm C}^{(1)}$  &  
model 2\footnotemark[$\ddagger$] & ${\Delta\,{\rm log}\,A_{\rm
 C}^{(2)}}$\footnotemark[$\S$] & 
log\,$A_{\rm C}^{(2)}$ & $\chi_{\rm BB}^{2}$  \\
 (1)   &   (2)  &  (3) & (4) & (5) & (6) & (7) & (8) & (9) & (10) \\
        \hline
 \endfirsthead
        \hline
obj. &  $T_{\rm eff}$ & log\,$g$ &  model 1\footnotemark[$*$] & ${\Delta\,{\rm
 log}\,A_{\rm C}^{(1)}}$\footnotemark[$\dagger$] & log\,$A_{\rm C}^{(1)}$  &  
model 2\footnotemark[$\ddagger$] & ${\Delta\,{\rm log}\,A_{\rm
 C}^{(2)}}$\footnotemark[$\S$] & 
log\,$A_{\rm C}^{(2)}$ & $\chi_{\rm BB}^{2}$  \\
 (1)   &   (2)  &  (3) & (4) & (5) & (6) & (7) & (8) & (9) & (10) \\
        \hline
\endhead
   \hline
\endfoot
   \hline
\multicolumn{9}{l}{\hbox to 0pt {\parbox{150mm}{\footnotesize
  \footnotemark[$*$] Model photosphere from the UCM grid.

  \footnotemark[$\dagger$] First correction to the assumed value for the models
   of the Cc series,  log\,$A_{\rm C}^{(0)}$ = -3.61.

  \footnotemark[$\ddagger$] Specified model for $T_{\rm eff}$ and log\,$g$
  in the second and third columns, respectively.

  \footnotemark[$\S$] Second correction to the value in the 6-th column,
    log\,$A_{\rm C}^{(1)}$.

}}}
\endlastfoot
   \hline
           &      &         &               &         &         &            &         &         &         \\
GJ~15A      & 3567 &   4.890 &    Cc3600c50  &   0.169 &  -3.441 & Ca3570c489 &  -0.055 &  -3.496 $\pm$ 0.099 &   4.381 \\
GJ~105B     & 3385 &   4.936 &    Cc3400c50  &   0.247 &  -3.363 & Ca3360c495 &  -0.030 &  -3.393 $\pm$ 0.067  &   2.820 \\
GJ~166C     & 3341 &   4.968 &    Cc3300c50  &   0.315 &  -3.295 & Ca3340c497 &   0.045 &  -3.250 $\pm$ 0.125 &   4.369 \\
GJ~176      & 3616 &   4.804 &    Cc3600c475 &   0.339 &  -3.271 & Ca3620c480 &   0.006 &  -3.265 $\pm$ 0.055 &   4.285 \\
GJ~179      & 3476 &   4.877 &    Cc3500c50  &   0.288 &  -3.322 & Ca3480c488 &  -0.041 &  -3.363 $\pm$ 0.064 &   3.188 \\
GJ~205      & 3801 &   4.710 &    Cc3800c475 &   0.594 &  -3.016 & Ca3800c471 &  -0.050 &  -3.066 $\pm$ 0.042  &   3.791 \\
GJ~212      & 3757 &   4.748 &    Cc3750c475 &   0.372 &  -3.238 & Ca3760c475 &  -0.019 &  -3.257 $\pm$ 0.087  &   3.330 \\
GJ~229      & 3707 &   4.766 &    Cc3700c475 &   0.428 &  -3.182 & Ca3710c477 &  -0.011 &  -3.193 $\pm$ 0.050  &   3.200 \\
GJ~231.1B   & 3442 &   4.897 &    Cc3400c50  &   0.111 &  -3.499 & Ca3440c490 &  -0.002 &  -3.501 $\pm$ 0.044  &   1.726 \\
GJ~250B     & 3567 &   4.827 &    Cc3550c475 &   0.214 &  -3.396 & Ca3570c483 &   0.032 &  -3.364 $\pm$ 0.079 &   3.234 \\
           &      &         &               &         &         &            &         &         &         \\
GJ~273-L    & 3150 &   5.040 &    Cc3200c50  &   0.101 &  -3.509 & Cc3150c504 &  -0.049 &  -3.558 $\pm$ 0.048  &   5.810 \\
GJ~273-H    & 3415 &   4.915 &    Cc3400c50  &   0.275 &  -3.335 & Ca3420c492 &  -0.003 &  -3.338 $\pm$ 0.054  &   4.959 \\
GJ~324B     & 3401 &   4.924 &    Cc3400c50  &   0.335 &  -3.275 & Ca3380c494 &  -0.017 &  -3.292 $\pm$ 0.064 &   5.100 \\
GJ~338A     & 3907 &   4.709 &    Cc3900c475 &   0.114 &  -3.496 & Ca3910c471 &  -0.039 &  -3.535 $\pm$ 0.037 &   2.613 \\
GJ~338B     & 3867 &   4.709 &    Cc3850c475 &   0.124 &  -3.486 & Ca3870c471 &  -0.042 &  -3.528 $\pm$ 0.043 &   2.043 \\
GJ~380      & 4081 &   4.643 &    Cc4100c475 &   0.470 &  -3.140 & Ca4080c464 &  -0.071 &  -3.211 $\pm$ 0.085 &   2.798 \\
GJ~406      & 2800 &   5.250 &    Cc2800c525 &   0.053 &  -3.557 & Cc2800c517 &  +0.010 &  -3.547 $\pm$ 0.058 &   5.558 \\
GJ~411      & 3465 &   4.857 &    Cc3450c475 &  -0.054 &  -3.655 & Cc3470c486 &   0.059 &  -3.605 $\pm$ 0.043 &   2.732 \\
GJ~412A     & 3497 &   4.843 &    Cc3500c475 &  -0.181 &  -3.791 & Cc3500c484 &   0.034 &  -3.757 $\pm$ 0.056 &   4.241 \\
GJ~436      & 3416 &   4.797 &    Cc3400c475 &   0.047 &  -3.563 & Cc3420c480 &   0.024 &  -3.539 $\pm$ 0.081 &   3.991 \\
           &      &         &               &         &         &            &         &         &         \\
GJ~526      & 3618 &   4.784 &    Cc3600c475 &   0.103 &  -3.507 & Cc3620c478 &   0.018 &  -3.489 $\pm$ 0.041 &   3.614 \\
GJ~581      & 3442 &   4.959 &    Cc3400c50  &   0.122 &  -3.488 & Ca3440c496 &   0.016 &  -3.472 $\pm$ 0.062 &   2.791 \\
GJ~611B    & 3202 &   5.063 &    Cc3200c50  &  -0.075 &  -3.685 & Cc3200c506 &  -0.008 &  -3.693  $\pm$ 0.075 &   4.315 \\
GJ~649      & 3660 &   4.784 &    Cc3650c475 &   0.131 &  -3.479 & Ca3660c478 &   0.005 &  -3.474 $\pm$ 0.053 &   1.268 \\
GJ~686      & 3538 &   4.842 &    Cc3550c475 &   0.184 &  -3.426 & Ca3540c484 &   0.019 &  -3.407 $\pm$ 0.077 &   3.591 \\
GJ~687     & 3413 &   4.811 &    Cc3400c475 &   0.259 &  -3.351 & Ca3410c481 &    0.019 &  -3.332 $\pm$ 0.055 &   5.592 \\
GJ~725A    & 3407 &   4.837 &    Cc3400c475 &   0.079 &  -3.531 & Cc3410c484 &   0.031 &  -3.500 $\pm$ 0.078 &   3.004 \\
GJ~725B-L    & 3104 &   4.790 &    Cc3100c475 &  -0.156 &  -3.766 & Cc3100c479 &   0.004 & -3.762 $\pm$ 0.098 &   4.218 \\
GJ~725B-H   & 3337 &   4.972 &    Cc3300c50  &   0.062 &  -3.548 & Cc3340c497 &   0.025 &  -3.523 $\pm$ 0.100 &  3.485 \\
GJ~768.1C  & 3470 &   4.880 &    Cc3500c50  &   0.249 &  -3.361 & Ca3470c488 &  -0.045 &  -3.406 $\pm$ 0.092 &   0.661 \\
           &      &         &               &         &         &            &         &         &         \\
GJ~777B     & 3310 &   4.991 &    Cc3300c50  &   0.348 &  -3.262 & Ca3310c499 &   0.038 &  -3.224 $\pm$ 0.075 &   0.827 \\
GJ~783.2B  & 3368 &   4.949 &    Cc3400c50  &   0.259 &  -3.351 & Ca3370c495 &  -0.007 &  -3.358 $\pm$ 0.073 &   4.048 \\
GJ~797B-NE  & 3473 &   4.878 &    Cc3500c50  &   0.138 &  -3.472 & Ca3470c488 &  -0.038 &  -3.510 $\pm$ 0.040 &   1.101 \\
GJ~797B-SW  & 3473 &   4.878 &    Cc3500c50  &   0.162 &  -3.448 & Ca3470c488 &  -0.026 &  -3.474 $\pm$ 0.034 &   1.076 \\
GJ~809      & 3692 &   4.720 &    Cc3700c475 &   0.159 &  -3.451 & Ca3690c472 &  -0.033 &  -3.484 $\pm$ 0.058 &   2.196 \\
GJ~820B     & 3932 &   4.679 &    Cc3950c475 &   0.200 &  -3.410 & Ca3930c468 &  -0.056 &  -3.466 $\pm$ 0.036 &   3.085 \\
GJ~849      & 3580 &   4.821 &    Cc3600c475 &   0.411 &  -3.199 & Ca3580c482 &   0.007 &  -3.192 $\pm$ 0.042 &   5.227 \\
GJ~876      & 3458 &   4.888 &    Cc3500c50  &   0.408 &  -3.202 & Ca3460c489 &  -0.045 &  -3.247 $\pm$ 0.045 &   2.499 \\
GJ~880      & 3713 &   4.716 &    Cc3710c475 &   0.397 &  -3.213 & Ca3710c472 &  -0.037 &  -3.250 $\pm$ 0.053 &   2.990 \\
GJ~884      & 3850 &   4.720 &    Cc3850c475 &   0.295 &  -3.315 & Ca3850c472 &  -0.046 &  -3.361 $\pm$ 0.049 &   2.153 \\
           &      &         &               &         &         &            &         &         &         \\
GJ~3348B    & 3448 &   4.893 &    Cc3400c50  &   0.270 &  -3.340 & Ca3480c488 &   0.021 &  -3.319 $\pm$ 0.088 &   0.563 \\
HIP~12961   & 3890 &   4.709 &    Cc3900c475 &   0.529 &  -3.081 & Ca3890c471 &  -0.038 &  -3.119 $\pm$ 0.077 &   2.233 \\
HIP~57050  & 3464 &   4.884 &    Cc3500c50  &   0.450 &  -3.160 & Ca3460c488 &  -0.047 &  -3.207 $\pm$ 0.062 &   3.808 \\
HIP~79431  & 3592 &   4.815 &    Cc3600c475 &   0.445 &  -3.165 & Ca3590c482 &   0.009 &  -3.156 $\pm$ 0.080 &   6.321 \\
           &      &         &               &         &         &            &         &         &         \\
\hline
\end{longtable}

\begin{table}[ht]
  \begin{center}
  \caption{Effect of the uncertainties in  the fundamental parameters
and metallicity on determination of the carbon  abundances   } 
\label{tab:8-th}
    \begin{tabular}{lcccccc}
      \hline
  obj.    & Ser. &   $T_{\rm eff}$   & log\,$g$  & 
$\xi_{\rm micro}$ & $\Delta\,{\rm log}\,A_{\rm C}$ & diff.\footnotemark[$*$]\\ 
(log\,$A_{\rm C}^{0}$) &   &  (K)  &  & (km\,sec$^{-1}$) &   &   \\
   \hline
  GJ\,338A & Cc  & 3900  & 4.75 & 1.0 & ~0.039  & ~0.000 \\
  (-3.535) & ---\footnotemark[$\dagger$] & 3850 & ---  & --- &~0.049 &~0.010 \\
           &  --- & 3950  & ---   &  ---  & ~0.032  & -0.007 \\
           &  ---  & 3900  & 4.50 &  ---  & -0.133 & -0.172 \\
           &  ---  &   ---   & 5.00 &  ---  & ~0.187  & ~0.148 \\
           &  ---  &   ---   & 4.75 & 0.5 & ~0.152  & ~0.113 \\
           &  ---  &   ---   &   ---  & 1.5 & -0.075 & -0.114 \\
           & Ca  &   ---   &   ---  & 1.0 & ~0.030 & -0.009 \\ 
\hline
  GJ\,436  &  Cc & 3400  & 4.75 & 1.0  & -0.024  & 0.000  \\
 (-3.539)  &  --- & 3350  &   ---  &  ---   & -0.051  & -0.027 \\
           &  ---  & 3450  &   ---  &  ---   & -0.000  & +0.024 \\
           &  ---  & 3400  & 4.50 &  ---   & -0.116  & -0.092 \\
           &  ---  &   ---   & 5.00 &  ---   &  0.045  & 0.069 \\
           &  ---  &   ---   & 4.75 &  0.5 &  0.075  & 0.099 \\
           &  ---  &   ---   &  ---   &  1.5 & -0.131  & -0.107 \\
           & Ca  &  ---    &  ---   &  1.0 & -0.023  & +0.001 \\
\hline   
    GJ\,406 & Cc &  2800 & 5.25  &  1.0 & 0.000    &  0.000 \\
   (-3.557) & ---  &  2750 &  ---    &  ---   & -0.067   & -0.067 \\      
            & ---  &  2850 &  ---    &  ---   & 0.069    &  0.069 \\    
            & ---  &  2800 & 5.00  &  ---   & 0.032    &  0.032  \\    
            & ---  &   ---   & 5.50  &  ---   & -0.031   & -0.031 \\     
            & ---  &   ---   & 5.25  & 0.5  & 0.051    & 0.051 \\     
            & ---  &   ---   &  ---    & 1.5  & -0.065   & -0.065 \\
            & Ca &   ---   &  ---    & 1.0  & 0.068    &  0.068 \\     
      \hline
   GJ\,412A & Cc &  3500 &  4.75 &  1.0  & -0.181  &  0.000  \\
    (-3.61) & Cm\footnotemark[$\ddagger$]  &  ---   &  ---  &   ---   & -0.275 
                &  -0.094  \\
     \hline     
  \multicolumn{6}{l}{\hbox to 0pt{\parbox{85mm}{\footnotesize
     \par\noindent
     \footnotemark[$*$]  Difference of $\Delta\,{\rm log}\,A_{\rm C}$ 
      against that for the reference model in the first line for each star.
    \par\noindent
    \footnotemark[$\dagger$] --- means identical with the same column
       of the preceding line.
   \par\noindent 
   \footnotemark[$\ddagger$] A metal poor model in which the abundances of
       all the metals (31 elements) except for He and Li are reduced by 
       0.20\,dex from the abundances of {\it case c}. 
       }}}
   \end{tabular}
  \end{center}
   \end{table}

\begin{longtable}{lccc}
  \caption{Carbon abundances in M dwarfs - a result by the BS method}
\label{tab:9-th}
     \hline
obj.   & ${\Delta\,{\rm log}\,A_{\rm C}^{(3)}}$\footnotemark[$*$] &
 log\,$A_{\rm C}^{(3)}$    & $\chi_{\rm BS}^{2}$  \\
    \hline
    \endfirsthead
        \hline
obj.   & ${\Delta\,{\rm log}\,A_{\rm C}^{(3)}}$\footnotemark[$*$] &
 log\,$A_{\rm C}^{(3)}$    & $\chi_{\rm BS}^{2}$  \\
   \hline
  \endhead
   \hline
\endfoot
   \hline
\multicolumn{4}{l}{\hbox to 0pt {\parbox{80mm}{\footnotesize
  \footnotemark[$*$]  Third correction to the value in the 9-th column
    of table\,8, log\,$A_{\rm C}^{(2)}$.

  \footnotemark[$\dagger$] The cases L and H differ in $T_{\rm eff}$ values 
 (see subsection\,3.1), and we finally adopt the carbon abundances of the 
   case H both  for GJ\,273 and GJ\,725B for the reason outlined in 
  subsection\,6.2.   

}}}
\endlastfoot
   \hline
        &         &                     &        \\
GJ15A  &  -0.105 &  -3.601 $\pm$ 0.107 &   3.660\\
GJ105B &  -0.075 &  -3.468 $\pm$ 0.061 &   2.788\\
GJ166C &  -0.116 &  -3.366 $\pm$ 0.135 &   3.812\\
GJ176  &  -0.087 &  -3.352 $\pm$ 0.066 &   3.660\\
GJ179  &  -0.064 &  -3.427 $\pm$ 0.114 &   3.346\\
GJ205  &  -0.050 &  -3.116 $\pm$ 0.075 &   3.728\\
GJ212  &  -0.039 &  -3.296 $\pm$ 0.109 &   3.196\\
GJ229  &  -0.078 &  -3.271 $\pm$ 0.074 &   3.113\\
GJ231.1B  &  -0.071 &  -3.572 $\pm$ 0.051 &   1.674\\
GJ250B  &  -0.049 &  -3.413 $\pm$ 0.104 &   3.054\\
        &         &                     &        \\
GJ273-L\footnotemark[$\dagger$]  &  -0.060 &  -3.618 $\pm$ 0.103 &   5.194\\
GJ273-H\footnotemark[$\dagger$]   &  -0.065 &  -3.403 $\pm$ 0.109 &   4.570\\
GJ324B   &  -0.068 &  -3.360 $\pm$ 0.126 &   4.872\\
GJ338A  &  -0.052 &  -3.587 $\pm$ 0.035 &   2.178\\
GJ338B  &  -0.050 &  -3.578 $\pm$ 0.035 &   1.681\\
GJ380  &  -0.073 &  -3.284 $\pm$ 0.062 &   2.828\\
GJ406  &  -0.065 &  -3.612 $\pm$ 0.098 &   4.735\\
GJ411  &  -0.066 &  -3.671 $\pm$ 0.055 &   1.981\\
GJ412A  &  -0.092 &  -3.849 $\pm$ 0.036 &   1.867\\
GJ436   &  -0.095 &  -3.634 $\pm$ 0.056 &   3.391\\
        &         &                     &        \\
GJ526   &  -0.064 &  -3.553 $\pm$ 0.038 &   2.765\\
GJ581   &  -0.088 &  -3.560 $\pm$ 0.049 &   2.410\\
GJ611B  &  -0.069 &  -3.762 $\pm$ 0.034 &   4.091\\
GJ649    &  -0.070 &  -3.544 $\pm$ 0.035 &   1.140\\
GJ686    &  -0.090 &  -3.497 $\pm$ 0.037 &   3.083\\
GJ687   &  -0.096 &  -3.428 $\pm$ 0.091 &   4.006\\
GJ725A  &  -0.084 &  -3.584 $\pm$ 0.092 &   2.394\\
GJ725B-L\footnotemark[$\dagger$]  &  -0.096 &  -3.858 $\pm$ 0.076 &   3.068\\
GJ725B-H\footnotemark[$\dagger$]  &  -0.082 &  -3.605 $\pm$ 0.079 &   3.109\\
GJ768.1C &  -0.098 &  -3.504 $\pm$ 0.077 &   0.573\\
        &         &                     &        \\
GJ777B   &  -0.015 &  -3.239 $\pm$ 0.155 &   0.827\\
GJ783.2B &  -0.055 &  -3.413 $\pm$ 0.095 &   3.815\\
GJ797B-NE  &  -0.025 &  -3.535 $\pm$ 0.087 &   1.086\\
GJ797B-SW  &  -0.032 &  -3.506 $\pm$ 0.086 &   1.018\\
GJ809    &  -0.066 &  -3.550 $\pm$ 0.044 &   2.128\\
GJ820B   &  -0.040 &  -3.506 $\pm$ 0.045 &   2.918\\
GJ849    &  -0.078 &  -3.270 $\pm$ 0.092 &   4.217\\
GJ876    &  -0.108 &  -3.355 $\pm$ 0.133 &   2.315\\
GJ880    &  -0.098 &  -3.348 $\pm$ 0.067 &   2.857\\
GJ884    &  -0.063 &  -3.424 $\pm$ 0.070 &   2.024\\
        &         &                     &        \\
GJ3348B  &  -0.081 &  -3.400 $\pm$ 0.106 &   0.502\\
HIP12961  &  -0.069 &  -3.188 $\pm$ 0.070 &   2.067\\
HIP57050 &  -0.052 &  -3.259 $\pm$ 0.122 &   3.419\\
HIP79431 &  -0.080 &  -3.236 $\pm$ 0.116 &   5.460\\
        &         &                     &        \\
\hline
\end{longtable}

\begin{table}
  \caption{Carbon abundances and metallicities in M dwarfs  } 
\label{tab:10-th}
  \begin{center}
    \begin{tabular}{lll}
      \hline
      \smallskip
   obj. &  [M/H]\footnotemark[$*$] & log\,$A_{\rm C}$\footnotemark[$\S$]\\ 
      \hline
     GJ~15A  &  -0.35 $\pm$ 0.2\footnotemark[$\dagger$] &  -3.60 $\pm$ 0.11  \\
     GJ~105B & -0.06 $\pm$ 0.15\footnotemark[$\ddagger$] & -3.47 $\pm$ 0.06 \\ 
     GJ~176  &  +0.04 $\pm$ 0.02\footnotemark[$\ddagger$] & -3.35 $\pm$ 0.07 \\
     GJ~205 &  +0.5~ $\pm$ 0.2\footnotemark[$\dagger$]  & -3.12 $\pm$ 0.08 \\
     GJ~229 &  +0.2~ $\pm$ 0.2\footnotemark[$\dagger$]  & -3.27 $\pm$ 0.07 \\
     GJ~250B & -0.05 $\pm$ 0.05\footnotemark[$\ddagger$] &  -3.41 $\pm$ 0.10 \\
     GJ~411  & -0.35 $\pm$  0.2\footnotemark[$\dagger$] & -3.67 $\pm$ 0.06 \\
     GJ~436  & +0.08 $\pm$ 0.05\footnotemark[$\ddagger$] & -3.63 $\pm$ 0.06 \\
     GJ~581  & -0.15 $\pm$ 0.03\footnotemark[$\ddagger$] & -3.56 $\pm$ 0.05 \\
     GJ~820B & -0.1~ $\pm$ 0.2\footnotemark[$\dagger$] & -3.51 $\pm$ 0.05 \\
     GJ~849  & +0.35 $\pm$ 0.10\footnotemark[$\ddagger$] & -3.27 $\pm$ 0.09 \\
     GJ~876  & +0.12 $\pm$ 0.15\footnotemark[$\ddagger$] & -3.36 $\pm$ 0.13 \\
     the Sun &      ~~0.0    &  -3.40 $\pm$ 0.05\footnotemark[$\|$]  \\
     the Sun &      ~~0.0    &  -3.61 $\pm$ 0.04\footnotemark[$\#$]   \\
      \hline
  \multicolumn{3}{l}{\hbox to 0pt{\parbox{85mm}{\footnotesize
     \par\noindent
     \footnotemark[$*$] [M/H] = log\,${ (A_{\rm M})_{*} }$-
         log\,${ (A_{\rm M})_{\odot} }$, where $A_{\rm M}$ is
          the number density of metal M relative to hydrogen.
     \par\noindent
     \footnotemark[$\dagger$] \citet{Mou78}. 
     \par\noindent
     \footnotemark[$\ddagger$]  \citet{One12}.
     \par\noindent
     \footnotemark[$\S$]  table\,10, this paper.
     \par\noindent
     \footnotemark[$\|$]  \citet{Gre91}.
      \par\noindent
     \footnotemark[$\#$]   \citet{All02}.
     }}}
   \end{tabular}
  \end{center}
   \end{table}

\begin{table}
  \caption{ Carbon and iron abundances in M dwarfs } 
\label{tab:11-th}
  \begin{center}
    \begin{tabular}{lcc}
      \hline
      \smallskip
   obj. & [Fe/H]\footnotemark[$*$] & log\,$A_{\rm C}$\footnotemark[$\dagger$]\\ 
      \hline
     GJ\,105B   &   -0.02  &  -3.47   \\
     GJ\,166C   &   0.08   &  -3.37    \\
     GJ\,176    &   -0.01  &  -3.35    \\
     GJ\,205    &   0.22   &  -3.12    \\
     GJ\,229    &   -0.01  &   -3.27  \\
     GJ\,250B   &   -0.10  &   -3.41  \\
     GJ\,273    &   -0.01  &   -3.40 \\
     GJ\,406    &   0.18   &   -3.61 \\
     GJ\,526    &  -0.20   &    -3.55 \\
     GJ\,581    &  -0.21   &    -3.56 \\
     GJ\,686    &  -0.37   &  -3.50  \\
     GJ\,849    &   0.24   &  -3.27  \\
     GJ\,876    &   0.15   &  -3.36  \\
     GJ\.880    &   0.07   &  -3.35   \\ 
     \hline
  \multicolumn{3}{l}{\hbox to 0pt{\parbox{85mm}{\footnotesize
     \par\noindent
     \footnotemark[$*$] \citet{Nev13}. 
     \par\noindent
     \footnotemark[$\dagger$] table\,10, this paper. 
     }}}
   \end{tabular}
  \end{center}
   \end{table}

\begin{table}
  \caption{New effective temperatures by the interferometry} 
\label{tab:12-th}
  \begin{center}
    \begin{tabular}{lccc}
      \hline
      \smallskip
  obj. & $T_{\rm eff}$\footnotemark[$*$] & $T_{\rm eff}$\footnotemark[$\dagger$] 
  &  difference     \\ 
      \hline
     GJ\,176    & 3679 $\pm$ 77  & 3616  & +63   \\
     GJ\,649    & 3590 $\pm$ 45  & 3660  & -70   \\ 
     GJ\,876    & 3129 $\pm$ 19  & 3458  & -329  \\
     \hline
  \multicolumn{4}{l}{\hbox to 0pt{\parbox{80mm}{\footnotesize
     \par\noindent
     \footnotemark[$*$] By interferometry \citep{Bra14}. 
     \par\noindent
     \footnotemark[$\dagger$] By $M_{\rm 3.4}$ - log\,$T_{\rm eff}$
     relation (table\,5, this paper).
     }}}
   \end{tabular}
   \end{center}
   \end{table}



{\small

\begin{longtable}{lccccccccccccc}
\caption{log\,$(W/\lambda)_{\rm obs}$\footnotemark[$*$] of CO blends (2-0 
band) in 42 M dwarfs\footnotemark[$\dagger$]--for electronic version only}
\label{tab:13-th}   
\hline
obj.      &   no.1  &   no.2  &   no.3  &   no.4  &   no.5  &   no.6  &   no.7  &   no.8  &   no.9  &   no.10 &   no.12 &   no.13 &   no.14 \\
   \hline  
\endfirsthead
   \hline
obj.      &   no.1  &   no.2  &   no.3  &   no.4  &   no.5  &   no.6  &   no.7  &   no.8  &   no.9  &   no.10 &   no.12 &   no.13 &   no.14 \\
   \hline   
\endhead
   \hline
\endfoot
   \hline
\multicolumn{14}{l}{\hbox to 0pt {\parbox{160mm}{\footnotesize
  \footnotemark[$*$] The equivalent widths are measured by referring to
 the pseudo-continuum.

\footnotemark[$\dagger$] The spectroscopic data for lines through no.1 to
 no.14  are given in table\,7 for corresponding Ref. nos.

}}}
\endlastfoot
   \hline
       &       &       &     &       &     &     &         &       &          &      &        &        &     \\
GJ15A     &  -4.836 &  -4.926 &  -4.834 &  -4.859 &  -4.855 &  -4.786 &  -4.766 &  -4.749 &  -4.817 &  -4.883 &  -4.887 &  -4.984 &  -4.882 \\
GJ105B    &  -4.756 &  -4.763 &  -4.710 &  -4.738 &  -4.780 &  -4.769 &  -4.718 &  -4.745 &  -4.770 &  -4.831 &  -4.851 &  -4.749 &  -4.676 \\
GJ166C    &  -4.750 &  -4.717 &  -4.653 &  -4.646 &  -4.774 &  -4.683 &  -4.632 &  -4.679 &  -4.704 &  -4.743 &  -4.881 &  -4.734 &  -4.705 \\
GJ176     &  -4.747 &  -4.735 &  -4.737 &  -4.757 &  -4.724 &  -4.712 &  -4.710 &  -4.750 &  -4.770 &  -4.723 &  -4.819 &  -4.789 &  -4.808 \\ 
GJ179     &  -4.751 &  -4.764 &  -4.815 &  -4.793 &  -4.803 &  -4.788 &  -4.734 &  -4.722 &  -4.753 &  -4.806 &  -4.776 &  -4.735 &  -4.730 \\
GJ205     &  -4.665 &  -4.665 &  -4.653 &  -4.687 &  -4.710 &  -4.652 &  -4.676 &  -4.698 &  -4.690 &  -4.692 &  -4.693 &  -4.698 &  -4.729 \\
GJ212     &  -4.734 &  -4.723 &  -4.804 &  -4.771 &  -4.718 &  -4.807 &  -4.712 &  -4.709 &  -4.706 &  -4.761 &  -4.740 &  -4.737 &  -4.754 \\
GJ229     &  -4.712 &  -4.718 &  -4.718 &  -4.747 &  -4.759 &  -4.736 &  -4.688 &  -4.707 &  -4.718 &  -4.713 &  -4.743 &  -4.767 &  -4.792 \\
GJ231.1B  &  -4.817 &  -4.809 &  -4.773 &  -4.803 &  -4.828 &  -4.851 &  -4.789 &  -4.783 &  -4.819 &  -4.847 &  -4.862 &  -4.788 &  -4.778 \\
GJ250B    &  -4.811 &  -4.786 &  -4.750 &  -4.786 &  -4.788 &  -4.868 &  -4.762 &  -4.745 &  -4.753 &  -4.740 &  -4.783 &  -4.768 &  -4.756 \\
       &       &       &     &       &     &     &         &       &          &      &        &        &     \\
GJ273     &  -4.763 &  -4.752 &  -4.727 &  -4.746 &  -4.814 &  -4.781 &  -4.724 &  -4.726 &  -4.744 &  -4.781 &  -4.760 &  -4.726 &  -4.714 \\
GJ324B    &  -4.750 &  -4.744 &  -4.710 &  -4.723 &  -4.748 &  -4.748 &  -4.724 &  -4.661 &  -4.724 &  -4.796 &  -4.758 &  -4.714 &  -4.737 \\
GJ338A    &  -4.810 &  -4.832 &  -4.804 &  -4.838 &  -4.822 &  -4.819 &  -4.785 &  -4.799 &  -4.817 &  -4.842 &  -4.862 &  -4.853 &  -4.934 \\
GJ338B    &  -4.811 &  -4.837 &  -4.803 &  -4.843 &  -4.824 &  -4.815 &  -4.780 &  -4.796 &  -4.818 &  -4.837 &  -4.865 &  -4.852 &  -4.919 \\
GJ380     &  -4.640 &  -4.710 &  -4.682 &  -4.703 &  -4.714 &  -4.711 &  -4.662 &  -4.679 &  -4.713 &  -4.687 &  -4.799 &  -4.716 &  -4.799 \\
GJ406     &  -4.546 &  -4.570 &  -4.517 &  -4.532 &  -4.561 &  -4.611 &  -4.537 &  -4.544 &  -4.545 &  -4.613 &  -4.622 &  -4.551 &  -4.547 \\
GJ411     &  -4.885 &  -4.833 &  -4.811 &  -4.830 &  -4.863 &  -4.881 &  -4.852 &  -4.859 &  -4.863 &  -4.922 &  -4.879 &  -4.848 &  -4.849 \\
GJ412A    &  -4.881 &  -4.896 &  -4.891 &  -4.917 &  -4.919 &  -4.901 &  -4.872 &  -4.913 &  -4.982 &  -4.939 &  -5.021 &  -4.968 &  -4.945 \\
GJ436     &  -4.808 &  -4.859 &  -4.768 &  -4.847 &  -4.825 &  -4.806 &  -4.756 &  -4.753 &  -4.802 &  -4.835 &  -4.910 &  -4.839 &  -4.825 \\
GJ526     &  -4.829 &  -4.823 &  -4.801 &  -4.832 &  -4.828 &  -4.811 &  -4.776 &  -4.816 &  -4.835 &  -4.814 &  -4.875 &  -4.872 &  -4.840 \\
       &       &       &     &       &     &     &         &       &          &      &        &        &     \\
GJ581     &  -4.773 &  -4.799 &  -4.763 &  -4.765 &  -4.824 &  -4.830 &  -4.776 &  -4.809 &  -4.836 &  -4.877 &  -4.889 &  -4.811 &  -4.818 \\
GJ611B    &  -4.886 &  -4.819 &  -4.744 &  -4.831 &  -4.813 &  -4.848 &  -4.820 &  -4.787 &  -4.807 &  -4.893 &  -4.925 &  -4.760 &  -4.738 \\
GJ649     &  -4.775 &  -4.803 &  -4.793 &  -4.817 &  -4.848 &  -4.778 &  -4.799 &  -4.821 &  -4.837 &  -4.810 &  -4.897 &  -4.850 &  -4.866 \\
GJ686     &  -4.815 &  -4.792 &  -4.700 &  -4.790 &  -4.759 &  -4.792 &  -4.770 &  -4.816 &  -4.775 &  -    &  -4.874 &  -4.803 &  -4.833 \\
GJ687     &  -4.754 &  -4.742 &  -4.688 &  -4.696 &  -4.761 &  -4.772 &  -4.712 &  -4.731 &  -4.765 &  -4.750 &  -4.778 &  -4.760 &  -4.744 \\
GJ725A    &  -4.800 &  -4.807 &  -4.736 &  -4.745 &  -4.899 &  -4.790 &  -4.774 &  -4.825 &  -4.837 &  -4.792 &  -4.838 &  -4.776 &  -4.791 \\
GJ725B    &  -4.806 &  -4.803 &  -4.720 &  -4.728 &  -4.851 &  -4.790 &  -4.785 &  -4.803 &  -4.832 &  -4.804 &  -4.967 &  -4.769 &  -4.761 \\
GJ768.1B  &  -4.751 &  -4.772 &  -4.719 &  -4.717 &  -4.837 &  -4.794 &  -4.744 &  -4.771 &  -4.821 &  -4.878 &  -4.865 &  -4.768 &  -4.805 \\
GJ777B    &  -4.699 &  -4.686 &  -4.664 &  -4.651 &  -4.719 &  -4.743 &  -4.618 &  -4.702 &  -4.735 &  -4.717 &  -4.752 &  -4.677 &  -4.571 \\
GJ783.2B  &  -4.786 &  -4.748 &  -4.690 &  -4.686 &  -4.786 &  -4.762 &  -4.723 &  -4.757 &  -4.782 &  -4.805 &  -4.811 &  -4.678 &  -4.725 \\
       &       &       &     &       &     &     &         &       &          &      &        &        &     \\
GJ797B-NE &  -4.854 &  -4.814 &  -4.780 &  -4.822 &  -4.836 &  -4.826 &  -4.807 &  -4.807 &  -4.837 &  -4.860 &  -4.824 &  -4.752 &  -4.725 \\
GJ797B-SW &  -4.844 &  -4.794 &  -4.778 &  -4.792 &  -4.827 &  -4.815 &  -4.807 &  -4.803 &  -4.818 &  -4.826 &  -4.814 &  -4.772 &  -4.709 \\
GJ809     &  -4.761 &  -4.788 &  -4.764 &  -4.801 &  -4.862 &  -4.817 &  -4.788 &  -4.797 &  -4.814 &  -4.867 &  -4.841 &  -4.848 &  -4.841 \\
GJ820B    &  -4.802 &  -4.799 &  -4.798 &  -4.794 &  -4.769 &  -4.790 &  -4.766 &  -4.772 &  -4.786 &  -4.788 &  -4.818 &  -4.849 &  -4.826 \\
GJ849     &  -4.743 &  -4.690 &  -4.702 &  -4.705 &  -4.714 &  -4.743 &  -4.697 &  -4.724 &  -4.718 &  -4.761 &  -4.745 &  -4.753 &  -4.785 \\ 
GJ876     &  -4.756 &  -4.748 &  -4.712 &  -4.710 &  -4.737 &  -4.727 &  -4.715 &  -4.688 &  -4.743 &  -4.745 &  -4.768 &  -4.863 &  -4.758 \\
GJ880     &  -4.697 &  -4.712 &  -4.732 &  -4.741 &  -4.746 &  -4.698 &  -4.728 &  -4.745 &  -4.759 &  -4.711 &  -4.794 &  -4.808 &  -4.817 \\
GJ884     &  -4.757 &  -4.744 &  -4.777 &  -4.740 &  -4.805 &  -4.727 &  -4.746 &  -4.753 &  -4.782 &  -4.797 &  -4.793 &  -4.862 &  -4.799 \\
GJ3348B   &  -4.766 &  -4.743 &  -4.710 &  -4.721 &  -4.755 &  -4.777 &  -4.688 &  -4.745 &  -4.765 &  -4.882 &  -4.793 &  -4.753 &  -4.727 \\
HIP12961  &  -4.719 &  -4.724 &  -4.665 &  -4.648 &  -4.689 &  -4.718 &  -4.649 &  -4.668 &  -4.695 &  -4.742 &  -4.724 &  -4.755 &  -4.789 \\
       &       &       &     &       &     &     &         &       &          &      &        &        &     \\
HIP57050  &  -4.737 &  -4.693 &  -4.654 &  -4.684 &  -4.707 &  -4.736 &  -4.701 &  -4.740 &  -4.730 &  -4.777 &  -4.745 &  -4.675 &  -4.706 \\
HIP79431  &  -4.692 &  -4.695 &  -4.703 &  -4.702 &  -4.698 &  -4.734 &  -4.654 &  -4.693 &  -4.719 &  -4.820 &  -4.730 &  -4.708 &  -4.766 \\
      &       &       &     &       &     &     &         &       &          &      &        &        &     \\
\hline
\end{longtable}
 
}

\end{document}